\documentclass[onecolumn,amsmath,amssymb,subfigure,12pt]{revtex4-1}
\usepackage{bm}
\usepackage{graphicx}
\usepackage[newitem,newenum]{paralist}
\usepackage{mathrsfs}
\usepackage{amssymb}
\usepackage{csquotes}
\usepackage{bm}
\usepackage[dvips]{color}
\usepackage{color}
\usepackage{fancybox}
\usepackage{fancybox}
\usepackage{fancyhdr}
\begin{document}
\title{Dissipation process of binary mixture gas in thermally relativistic flow}
\author{Ryosuke Yano}
\affiliation{Department of Advanced Energy, University of Tokyo, 5-1-5 Kashiwanoha, Kashiwa, Chiba 277-8561, Japan}
\email{yano@k.u-tokyo.ac.jp}
\begin{abstract}
In this paper, we discuss dissipation process of the binary mixture gas in the thermally relativistic flow \textcolor{black}{by focusing on the characteristics of the diffusion flux}. As an analytical object, we consider the relativistic rarefied-shock layer problem around the triangle prism. Numerical results of the diffusion flux are compared with the Navier-Stokes-Fourier (NSF) order approximation of the diffusion flux, which is calculated using the diffusion and thermal-diffusion coefficients by Kox \textit{et al}. [Physica A, 84, 1, pp.165-174 (1976)]. In the case of the uniform flow with the small Lorentz contraction, the diffusion flux, which is obtained by calculating the relativistic Boltzmann equation, is roughly approximated by the NSF order approximation inside the shock wave, whereas the diffusion flux in the vicinity of the wall is markedly different from the NSF order approximation. The magnitude of the diffusion flux, which is obtained by calculating the relativistic Boltzmann equation, is similar to that of the NSF order approximation inside the shock wave, unlike the pressure deviator, dynamic pressure and heat flux, even when the Lorentz contraction in the uniform flow becomes large, because the diffusion flux does not depend on the generic Knudsen number from its definition in Eckart's frame. \textcolor{black}{Finally, the author concludes that the accurate diffusion flux must be calculated from the particle four flow, which is formulated using the four velocity distinguished by each species of particles.}
\end{abstract}

\maketitle

\section{Introduction}
The relativistic hydrodynamics has been a significant issue for understanding of the quark gluon plasma (QGP) \cite{Hatsuda} in Relativistic Heavy Ion Collider (RHIC) \cite{RHIC} and Large Hadron Collider (LHC) \cite{LHC} or astrophysical phenomena such as space jet \cite{Mizuta}. In particular, dissipation process of the relativistic matter has been discussed in the \textcolor{black}{framework} of the relativistic kinetic theory, in which various types of the relativistic hydrodynamic equation have been discussed from the viewpoint of the rational mechanics \cite{Muller} \cite{Denicol}. The kinetic analyses of dissipation process of the rarefied relativistic flow have been done by calculating the relativistic Boltzmann equation (RBE). For example, Bouras \textit{et al}. calculated the Riemann problem \cite{Bouras1} or the Mach cone in the QGP jet \cite{Bouras2} by calculating the RBE, whereas Yano \textit{et al}. \cite{Yano1} calculated the rarefied shock layer problem by calculating the RBE to investigate two types of relativistic effects, namely, thermally relativistic effect and Lorentz contraction effect, on dissipation process, where thermally relativistic matter is characterized using the thermally relativistic measure $\chi$ ($\chi=mc^2/k\theta$: $m$: mass of a \textcolor{black}{particle}, $c$: speed of light, $k$: Boltzmann constant, $\theta$: temperature) such as $\chi \le 100$. In the author's previous studies \cite{Yano1} \cite{Yano2}, the composition of thermally relativistic matter is limited to hard spherical particles with the equal mass and diameter, \textcolor{black}{which never reflect the realistic collisional cross section of the QGP \cite{Hatsuda} \cite{Peskin}}. Numerical results of dissipating terms such as the dynamic pressure and heat flux for the single component gas were compared with analytical results, which are obtained using the Navier-Stokes-Fourier (NSF) or Burnett order approximation \cite{Yano1} \cite{Yano2}, where we applied transport coefficients for single component hard spherical particles such as the bulk viscosity, viscosity coefficient and thermal conductivity, which were calculated by Groot \textit{et al}. \cite{Groot} or Cercignani and Kremer \cite{Cercignani}.\\
In recent studies of the QGP, the characteristics of the thermally relativistic mixture gas were discussed through the formulation of the viscosity coefficients of binary mixture of partons by Itakura \textit{et al}. \cite{Itakura} \cite{memo2} or El \textit{et al}. \cite{El}. Of course, the most classical formulation of transport coefficients of the thermally relativistic multi-component gas was obtained by van Leeuwen \textit{et al}. \cite{Leeuwen}. In the recent study by Wiranata \textit{et al}. \cite{Wiranata}, they calculated the viscosity coefficient of the thermally relativistic binary mixtures of \textcolor{black}{hard spherical particles} on the basis of its classical formulation by van Leeuwen \textit{et al}. \cite{Leeuwen}. On the other hand, we know that the diffusion flux is a markedly significant physical quantity from past studies of the nonrelativistic mixture fluids by Onsager \cite{Onsager}, Meixner \cite{Meixner}, Truesdell \cite{Truesdell} and M$\ddot{\mbox{u}}$ller-Ruggeri \cite{Muller}, \textcolor{black}{when we discuss the characteristics of the mixture fluids.}\\
\textcolor{black}{We, however, can not say that we understood the full characteristics of the diffusion flux in the relativistic regime, because the particle four flow \cite{Leeuwen} \cite{Kremer} was defined using the averaged four velocity over all the species of particles and diffusion flux instead of the four velocity distinguished by each species of particles, so that the diffusion flux was included in the particle four flow of the species \enquote{$a$} as a dissipating term such as the dynamic pressure, pressure deviator and heat flux. Consequently, the diffusion flux was expressed with gradients of the five field variables (the density, flow velocity and temperature) \cite{Cercignani} as a result of Chapman-Enskog method \cite{Leeuwen} or the first Maxwellian iteration of Grad's moment equations \cite{Kremer}. Indeed, the accurate diffusion flux must be calculated from the particle four flow by solving the RBE, which postulates the four velocity distinguished by each species of particles, because the diffusion flux is not a dissipating term, which depends on generic Knudsen number \cite{Yano2}, unlike the dynamic pressure, pressure deviator and heat flux, when we define the diffusion flux using the four velocity distinguished by each species of particles. Thus, we make it our primary aim to investigate the characteristics of the diffusion flux of the thermally relativistic gas by solving the RBE, numerically. As a supplemental study to attain our aim, we investigate whether the NSF order approximation of the diffusion flux by Kox \textit{et al}. \cite{Kox} demonstrates the diffusion flux, which is obtained by solving the RBE, with a good accuracy, when the binary mixture gas is composed of two species of hard spherical particles with equal masses and different diameters. The diffusion flux must be calculated from the RBE using the averaged four velocity, which is defined by four velocities of all the species of particles. As far as the author knows, such an averaged four velocity, which is defined by four velocities of all the species of particles, has not been discussed in the previous studies, explicitly.\\  
As an additional object of this study, we consider effects of diffusive terms in Grad's (Marle's \cite{Marle}) $28$ moment equations for the binary mixture gas, which is composed of two species of hard spherical particles with equal masses and different diameters, when we express Grad's $28$ moment equations with the flow velocity and temperature, which are defined for each species of particles. Finally, differences in Grad's 14 moments between different species of particles are included in the NSF law of the dynamic pressure, pressure deviator and heat flux. The investigation of the characteristics of the diffusion flux and effects of diffusive terms in the NSF law in the thermally relativistic mixture gas attributes to understanding of dissipation process of the thermally relativistic mixture gas.\\
To investigate the accuracy of the NSF order approximation of the diffusion flux by Kox \textit{et al}., and effects of diffusive terms in the NSF law, we calculate the thermally relativistic rarefied-shock layer around the triangle prism using the RBE. As a solver of the RBE, we use the direct simulation Monte Carlo (DSMC) method \cite{Bird} \cite{Yano1}. The shock layer problem is suitable to investigate the characteristics of dissipating terms such as the diffusion flux, dynamic pressure, pressure deviator, and heat flux \cite{Yano1}, because it includes the thermally nonequilibrium regime such as the steady shock wave and thermal boundary layer, in which dissipating terms are expected to be nonzero values.\\}
This paper is organized as follows. In Sec. II, we calculate Grad's 28 moment equations for the binary mixture gas (species $A$ and $B$) to understand diffusive terms in the NSF law, and define the NSF law for the dynamic pressure, pressure deviator and heat flux by taking the first Maxwellian iteration of Grad's 28 moment equations and neglecting all the diffusive terms in the NSF law, when masses of two species of hard spherical particles are equal. Next, we review the NSF law for the diffusion flux by Kox \textit{et al}. The NSF law defined in Sec. II is necessary to calculate NSF order approximations of the diffusion flux, dynamic pressure, pressure deviator and heat flux in Sec. III using numerical datum of five field variables \cite{Cercignani}. In Sec. III, we calculate the thermally relativistic rarefied-shock layer, which is constituted of the binary mixture gas, by solving the RBE. Numerical results of dissipating terms such as the diffusion flux, dynamic pressure, pressure deviator and heat flux, are compared with their analytical results (NSF order approximations), which are calculated using transport coefficients defined in Sec II. Additionally, we mention to effects of the mass ratio in the binary mixture gas on dissipation process, briefly. Finally, we make concluding remarks in Sec. IV.
\section{RBE for mixture gas and NSF order approximations}
\subsection{RBE for mixture gas, balance equations and definition of diffusion flux}
Firstly, the RBE for the mixture gas is written as
\begin{eqnarray}
p_a^\alpha \partial_\alpha f_a\left(\bm{p}_a\right)=\sum_b \int_{\mathscr{P}_b^3} \int_{\Omega} \left[f_a\left(\bm{p}_a^\prime\right) f_b\left(\bm{p}_b^\prime\right)-f_a\left(\bm{p}_a\right)f_b\left(\bm{p}_b\right)\right] F_{ab} \sigma_{ab} d\Omega \frac{d^3\bm{p}_b}{p^0_b},
\end{eqnarray}
where subscripts \enquote{$a$} and \enquote{$b$} correspond to the species of particles, respectively, $p_a^\alpha=m_a \gamma\left(v_a\right) \left(c,v_a^i\right)$ and $p_b^\alpha=m_b \gamma\left(v_b\right) \left(c,v_b^i\right)$ ($i=1,2,3$) are four momentums of two colliding species \enquote{$a$} and \enquote{$b$}, in which $\gamma\left(v_a\right)$ and $\gamma\left(v_b\right)$ are Lorentz factor of species \enquote{$a$} and \enquote{$b$}. In Eq. (1), $\bm{p}_a^\prime$ and $\bm{p}_b^\prime$ are momentum vectors after the binary collision. In Eq. (1), $f_a\left(\bm{p}_a\right)\equiv f_a\left(t,x^i,\bm{p}_a\right)$ ($f_b\left(\bm{p}_b\right)$) is the distribution function of the species $a$ ($b$). $\sigma_{ab}$ is the differential cross section between species \enquote{$a$} and \enquote{$b$}. In Eq. (1), $F_{ab}=\sqrt{\left(p^a_\alpha p_b^\alpha\right)^2-m_a^2 m_b^2c^4}$ is the Lorentz invariant flux and $\Omega$ is the solid angle on the spherical surface with the radius $1$. In Eq. (1), $\mathscr{P}_b^3=(0 \le \left|\bm{p}_b\right| <\infty)$ is the momentum space of the species \enquote{$b$}.\\
Multiplying $\psi\left(\bm{p}_a\right)=p_a^\beta p_a^\gamma p_a^\delta...$ by both sides of Eq. (1) and integrating over $d^3 \bm{p}_a/p_a^0$, we obtain
\begin{eqnarray}
&&\int_{\mathscr{P}_a} \psi\left(\bm{p}_a\right) p_a^\alpha \partial_\alpha f_a \frac{d^3\bm{p}_a}{p_a^0}=\sum_b \int_{\Omega}\int_{\mathscr{P}_a\times\mathscr{P}_b}\left[\psi\left(\bm{p}_a^\prime\right)-\psi\left(\bm{p}_a\right)\right] f_a\left(\bm{p}_a\right) f_b\left(\bm{p}_b\right) F_{ab} \sigma_{ab} d\Omega \frac{d^3 \bm{p}_b}{p^0_b}\frac{d^3 \bm{p}_a}{p^0_a}\nonumber \\
&=& \sum_b \Psi_{ab}^{\beta \gamma \delta...},
\end{eqnarray}
Substituting $\psi\left(\bm{p}_a\right)=c$, $cp_a^\beta$ and $cp_a^\beta p_a^\gamma$ into Eq. (2), we obtain
\begin{eqnarray}
\partial_\alpha N_a^\alpha&=&0,\\
\partial_\alpha T_a^{\alpha\beta}&=&\sum_b \Psi_{ab}^\beta=\Psi_a^\beta,\\
\partial_\alpha T_a^{\alpha\beta\gamma}&=&\sum_b \Psi_{ab}^{\beta\gamma}= \Psi_a^{\beta\gamma},
\end{eqnarray}
\textcolor{black}{$N_a^\alpha$ is the particle four flow of the species \enquote{$a$} and $T_a^{\alpha\beta}$ is the energy-momentum tensor of the species \enquote{$a$}.}
From the symmetry between species \enquote{$a$} and \enquote{$b$} in Eq. (2), we readily obtain
\begin{eqnarray}
&&\sum_a \int_{\mathscr{P}_a} \psi\left(\bm{p}_a\right) p_a^\alpha \partial_\alpha f_a \frac{d^3\bm{p}_a}{p_a^0}=\frac{1}{2} \sum_a \sum_b \int_{\Omega}\int_{\mathscr{P}_a\times \mathscr{P}_b}\left[\psi\left(\bm{p}_a^\prime\right)+\psi\left(\bm{p}_b^\prime\right)-\psi\left(\bm{p}_a\right)-\psi\left(\bm{p}_b\right)\right] \nonumber \\
&& f_a\left(\bm{p}_a\right) f_b\left(\bm{p}_b\right) F_{ab} \sigma_{ab}d\Omega \frac{d^3 \bm{p}_b}{p^0_b}\frac{d^3 \bm{p}_a}{p^0_a}.
\end{eqnarray}
From the mass and momentum-energy conservation, namely, ${p^\alpha_a}^\prime+{p^\alpha_b}^\prime=p^\alpha_a+p^\alpha_b$, we obtain following relation by substituting $\psi=c$, $cp_a^{\beta}$ and $cp_a^\beta p_a^\gamma$ into Eq. (6)
\begin{eqnarray}
\partial_\alpha \sum_a N_a^\alpha&=&\partial_\alpha N^\alpha=0,\\
\partial_\alpha \sum_a T_a^{\alpha\beta}&=&\partial_\alpha T^{\alpha\beta}=0,\\
\partial_\alpha \sum_a T_a^{\alpha\beta\gamma}&=&\partial_\alpha T^{\alpha\beta\gamma}=\Psi^{\beta\gamma},
\end{eqnarray}
where $\sum_a N_a^\alpha=N^\alpha$, $\sum_a T_a^{\alpha\beta}=T^{\alpha\beta}$, $\sum_a \Psi^\beta_a=0$ and $\sum_a \Psi_a^{\beta\gamma}=\Psi^{\beta\gamma}$.\\
In Eckart's frame \cite{Eckart}, $N_a^\alpha$ is written as
\begin{eqnarray}
N_a^\alpha&=&n_a U_a^\alpha,\nonumber \\
&=&n_a \bar{U}^\alpha+J_{a}^\alpha,
\end{eqnarray}
where $n_a$ is the number density of the species \enquote{$a$}, \textcolor{black}{$U_a^\alpha\equiv \gamma\left(u_a\right)\left(c,u_a^i\right)$ is the four velocity of the species \enquote{$a$}} ($u_a^i$: flow velocity of the species \enquote{$a$}), $J_a^\alpha$ is the diffusion flux of the species \enquote{$a$}, and $\bar{U}^\alpha$ is the averaged four-velocity \textcolor{black}{over all the species}. In our discussion, we consider projected moments in Eckart's frames, which are defined for each species, respectively. Meanwhile, readers remind that previous studies \cite{Kox} \cite{Kremer} used the common Eckart's frame for all the species using not $U_a^\alpha$ but $\bar{U}^\alpha$, \textcolor{black}{where $J_{a}^\alpha$ was regarded as a dissipating term, which is formulated using the five field variables as well as the dynamic pressure, pressure deviator and heat flux}. \textcolor{black}{In later numerical analyses of the RBE on the basis of the DSMC method, we calculate $N_a^\alpha=c \int_{\mathscr{P}^3} f_a d^3 \bm{p}_a/p_a^0$ for the species \enquote{$a$}, from which $J_a^\alpha$ is calculated using $\bar{U}^\alpha$. Therefore, we must define $\bar{U}^\alpha$ using ${U}_a^\alpha$, because the definition of $\bar{U}^\alpha$ with ${U}_a^\alpha$ has not been discussed in the previous studies, as far as the author knows.}\\
Summing over all the species in both sides of Eq. (10), we obtain
\begin{eqnarray}
\sum_a n_a U_a^\alpha=\sum_a n_a \bar{U}^\alpha=n \bar{U}^\alpha,
\end{eqnarray}
where $n=\sum_a n_a$ and $\sum_a J_a=0$ \textcolor{black}{are} used.\\
From Eq. (11), the averaged flow velocity ($\bar{u}^i$) and four-velocity are obtained using the relation $n \bar{U}^0=\sum_a n_a U_a^0$ in Eq. (11), as follows
\begin{eqnarray}
\bar{u}^i&=&\frac{\sum_a n_a U_{\textcolor{black}{a}}^i}{\sum_a n_a \gamma(u_a)},\nonumber \\
\bar{U}^\alpha&=&\gamma\left(\bar{u}\right)\left(c,\bar{u}^i\right).
\end{eqnarray}
\textcolor{black}{In later numerical analyses of the RBE on the basis of the DSMC method, we can also use the simple relation $\bar{U}^\alpha=\sum_a N_a^\alpha /n$ from Eq. (11).}\\
Using $\bar{U}^\alpha$ in Eq. (12), the diffusion flux ($J_a^\alpha$) is obtained as \cite{Kremer}
\begin{eqnarray}
J_a^\alpha=\bar{\Delta}_{\beta}^\alpha{N_a}^\beta,
\end{eqnarray}
where $\bar{\Delta}_{\beta}^\alpha=\bar{\Delta}_{\beta\gamma}\bar{\Delta}^{\alpha\gamma}$ and $\bar{\Delta}^{\alpha\beta}\equiv\eta^{\alpha\beta}-\bar{U}^\alpha \bar{U}^\beta/c^2$, in which $\eta^{\alpha\beta}=\mbox{diag}\left(+1,-1,-1,-1\right)$. Of course, $J_a^\alpha=0$ in Eq. (13), when we use the relation $\Delta_{\alpha\beta}\equiv\eta^{\alpha\beta}-U_a^\alpha U_a^\beta/c^2$ instead of $\bar{\Delta}_{\alpha\beta}$. Thus, $J_a^\alpha$ never emerges, when we use the four-velocity $U_a^\alpha$ to define $N_a^\alpha$, $T_a^{\alpha\beta}$ and $T_a^{\alpha\beta\gamma}$. Consequently, $f_a$ can be expanded using Grad's 14 moment equation as \cite{Cercignani}
\begin{eqnarray}
f_a \sim \left[f_a\right]_{14}=f_a^{MJ}\left(n_a,\theta_a,\bm{u}_a\right)\left(1+\frac{\Pi_a}{p_a}\mathcal{A}_a+\frac{{q_a}_\alpha}{p_a}\mathcal{B}_a^\alpha+\frac{{\Pi_a}_{\left<\alpha\beta\right>}}{p_a} \mathcal{C}^{\alpha\beta}_a \right),
\end{eqnarray}
where \textcolor{black}{$\theta_a$ is the temperature of the species \enquote{$a$}, $\Pi_a$ is the dynamic pressure of the species \enquote{$a$}, $q_a^\alpha$ is the heat flux of the species \enquote{$a$}, ${\Pi_a}_{\left<\alpha\beta\right>}$ is the pressure deviator of the species \enquote{$a$}}, $f_a^{MJ}\left(n_a,\theta_a,\bm{u}_a\right)=n_a/(4\pi m_a^2 ck \theta_a K_2(\chi_a)) \exp\left[-U_a^\alpha {p_a}_\alpha/(k\theta_a)\right]$ is the Maxwell-J$\ddot{\mbox{u}}$ttner function \textcolor{black}{of} the species \enquote{$a$} \cite{Cercignani}, in which $K_n$ is the n-th order modified Bessel function of the second kind, $\chi_a=m_a c^2/\left(k \theta_a\right)$, and $p_a=n_a k\theta_a$ is the static pressure \textcolor{black}{of} the species \enquote{$a$}. $\mathcal{A}_a$, $\mathcal{B}_a^\alpha$ and $\mathcal{C}^{\alpha\beta}_a$ in Eq. (14) are defined as \cite{Cercignani}
\begin{eqnarray}
\mathcal{A}_a&=&\frac{1-5 G_a\chi_a-\chi_a^2+G_a^2 \chi_a^2}{20 G_a+3\chi_a-13 G_a^2 \chi_a-2G_a \chi_a^2+2 G_a^3 \chi_a^2} \nonumber \\
&& \times \left[\frac{15 G_a+2\chi_a-6 G_a^2 \chi_a+5 G_a \chi_a^2+\chi_a^3-G_a^2\chi_a^3}{1-5 G_a\chi_a-\chi_a^2+G_a^2 \chi_a^2}\right. \nonumber \\
&& \left. +\frac{3\chi_a}{m_a c^2}\frac{6 G_a+\chi_a-G_a^2 \chi_a}{1-5 G_a\chi_a-\chi_a^2+G_a^2 \chi_a^2} {U_a}_\alpha p_a^\alpha+\frac{\chi_a}{m_a^2 c^4} {U_a}_\alpha {U_a}_\beta p_a^\alpha p_a^\beta \right], \nonumber \\
\mathcal{B}_a^\alpha&=&\frac{\chi_a}{\chi_a+5 G_a-G_a^2 \chi_a}\left[\frac{G_a}{m_a c^2}p_a^\alpha-\frac{1}{m_a^2 c^4}{U_a}_\beta p_a^\alpha p_a^\beta \right], \nonumber \\
\mathcal{C}_a^{\alpha\beta}&=&\frac{\chi_a}{2 G_a} \frac{1}{m_a^2 c^2} p_a^\alpha p_a^\beta,
\end{eqnarray}
where $G_a\equiv K_3(\chi_a)/K_2(\chi_a)$.\\
Substituting $f_a=\left[f_a\right]_{14}$ into Eqs. (3)-(5), we can evaluate balance equations of $N^\alpha_a$, $T_a^{\alpha\beta}$ and $T_a^{\alpha\beta\gamma}$ using Grad's 14 moments, as discussed in Sec. II-(B). Here, we find that the left hand sides of balance equations in Eqs. (3)-(5) are quite same as those for the single component gas. Therefore, effects by the mixture gas appear in right hand sides of Eqs. (4) and (5), exclusively. The mathematical difficulties exist in the calculation of such right hand sides of Eqs. (4) and (5), so that previous studies are limited to hard spherical particles \textcolor{black}{with} $m_a=m_b=m$ \cite{Kox} or Israel-Stewart particles \textcolor{black}{with} $m_a \sim m_b$ \cite{Kremer}. In this paper, we will not try to complete the calculation of $\Psi_{ab}^\beta$ and $\Psi_{ab}^{\beta\gamma}$ in Eqs. (4) and (5) beyond previous studies.
\subsection{Grad's 28 moment equations and NSF law for $\Pi_a$, $\Pi_a^{\left<\alpha\beta\right>}$ and $q_a^\alpha$}
In this subsection, we calculate the NSF law for $\Pi_a$, $\Pi_a^{\alpha\beta}$ and $q_a^\alpha$ using Grad's 28 moment equations for two species (\enquote{$a$}$=A$ and $B$), \textcolor{black}{where $m_A=m_B=m$ and $d_A \neq d_B$ ($d_a$; diameter of the species \enquote{$a$}) are assumed}. Firstly, we calculate Grad's 28 moment equations for the binary mixture gas using Eq. (14) in Eckart's frame \cite{Eckart}, from which we calculate the NSF law by taking the first Maxwellian iteration \cite{Kremer}.\\
In Eckart's frame, $N_a^\alpha$, $T_a^{\alpha\beta}$ and $T_a^{\alpha\beta\gamma}$ in left hand sides of Eqs. (3)-(5) are obtained using Grad's 14 moment equation in Eq. (14), namely, substituting $f=\left[f_a\right]_{14}$ into their definitions as \cite{Cercignani}
\begin{eqnarray}
N_a^\alpha &=&n_a U_a^\alpha,\\
T_a^{\alpha \beta}&=&\Pi_a^{\left<\alpha \beta\right>}-\left(p_a+\Pi_a\right)\Delta^{\alpha\beta}+\frac{1}{c^2}\left(U_a^\alpha q_a^\beta+U_a^\beta q_a^\alpha\right)+\frac{e_a n_a}{c^2}U_a^\alpha U_a^\beta,\\
T_a^{\alpha\beta\gamma}&=&\left(n_a C_1+C_2 \Pi_a\right)U_a^\alpha U_a^\beta U_a^\gamma+\frac{c^2}{6}\left(n_a m^2-n_a C_1-C_2\Pi_a\right)\left(\eta^{\alpha\beta}U_a^\gamma+\eta^{\alpha\gamma}U_a^\beta \right. \nonumber \\
&&\left.+\eta^{\beta\gamma}U_a^\alpha\right)+C_3\left(\eta^{\alpha \beta}q_a^\gamma+\eta^{\alpha\gamma}q_a^\beta+\eta^{\beta \gamma} q_a^\alpha \right)-\frac{6}{c^2}C_3\left(U_a^\alpha U_a^\beta q_a^\gamma+U_a^\alpha U_a^\gamma q_a^\beta\right.\nonumber \\
&& \left. +U_a^\beta U_a^\gamma q_a^\alpha \right)+C_4\left(\Pi_a^{\left<\alpha\beta\right>}U_a^\gamma+\Pi_a^{\left<\alpha\gamma\right>}U_a^\beta+\Pi_a^{\left<\beta\gamma\right>}U_a^\alpha\right),
\end{eqnarray}
where $C_1=\frac{m_a^2}{\chi_a}\left(\chi_a+6G_a\right)$,\\ $C_2=-\frac{6 m_a}{c^2\chi_a}\left[2\chi_a^3-5\chi_a+(19\chi_a^2-30)G_a-(2\chi_a^3-45\chi_a)G_a^2-9\chi_a^2G_a^3\right]$\\
$\times \left(20G_a+3\chi_a-13G_a^2\chi_a -2\chi_a^2G_a+2\chi_a^2G_a^3\right)^{-1}$, $C_3=-\frac{m_a}{\chi_a}\left(\chi_a+6G_a-G_a^2\chi_a\right)\left(\chi_a+5G_a-G_a^2\chi_a\right)^{-1}$ and $C_4=m_a\left(G_a\chi_a\right)^{-1}\left(\chi_a+6G_a\right)$. In Eq. (17), $e_a$ is the energy density \textcolor{black}{of} the species \enquote{$a$}.\\
Provided that we restrict ourselves to $m_A=m_B=m$ in the binary mixture gas, differences of the right hand side of Eq. (2) from that in the single component gas are forms of $\sigma_{ab}$ and $f_b$ because of $d_a \neq d_b$ and $f_a \neq f_b$. \textcolor{black}{In other words, the difference between $f_a$ and $f_b$ can be demonstrated by differences of Grad's 14 moments, when $f_a$ and $f_b$ are approximated by $\left[f_a\right]_{14}$ and $\left[f_b\right]_{14}$, respectively.} As a result of Eqs. (3)-(5) and (16)-(18), we obtain moment equations of $\Pi_a$, $\Pi^{\left<\alpha\beta\right>}_a$ and $q_a^\alpha$ by multiplying ${U_a}_\alpha {U_a}_\beta/c^4$, $\Delta_\alpha^\gamma \Delta_\beta^\delta-\Delta_{\alpha\beta}\Delta^{\gamma\delta}/3$ and $-\Delta_\alpha^\gamma {U_a}_\beta$ by both sides of Eq. (5) and neglecting nonlinear terms as
\begin{eqnarray}
&&\frac{C_2}{2} D \Pi_a+\frac{1}{2}\left(m^2+C_1\right) D n_a-\frac{\chi_a}{2 \theta_a} n_a C_1^\prime D \theta_a-5 \frac{C_3}{c^2} \nabla^\alpha {q_a}_\alpha \nonumber \\
&&+\frac{1}{6}\left(n_a m^2+5 n_a C_1\right)\nabla^\alpha {U_a}_\alpha=\sum_b \frac{1}{c^4} \Psi_{ab}^{\alpha\beta} {U_a}_\alpha {U_a}_\beta=-\frac{3}{c^2} \sum_b \mathfrak{B}_{ab} \left(\Pi_a+\delta \Pi_{ab} \right),\\
&&C_4 D \Pi^{\left<\alpha \beta\right>}_a+2 C_3 \nabla^{<\alpha} q_a^{\beta>}+\frac{c^2}{3}\left(n_a m^2-n_a C_1\right) \nabla^{<\alpha} U_a^{\beta>} \nonumber \\
&&=\sum_b \left(\Delta_\alpha^\gamma \Delta_\beta^\delta-\frac{1}{3}\Delta_{\alpha\beta}\Delta^{\gamma\delta}\right)\Psi_{ab}^{\left<\gamma \delta\right>}=-\sum_b \mathfrak{C}_{ab} \left(\Pi_a^{\left<\alpha\beta\right>}+\delta \Pi_{ab}^{\left<\alpha \beta\right>}\right),\\
&&5 C_3 D q_a^\alpha-\frac{c^4}{6}\left[\left(m^2-C_1\right)\nabla^\alpha n_a+\frac{\chi_a}{\theta_a} n_a C_1^\prime \nabla^\alpha \theta_a-C_2 \nabla^\alpha \Pi \right]-c^2 C_4 \nabla_\beta \Pi_a^{\left<\alpha \beta\right>} \nonumber\\
&&-\frac{c^2}{6} \left(n_a m^2+5 n_a C_1\right) D U_a^\alpha=-\sum_b \Psi_{ab}^{\beta \gamma} 
{U_a}_\beta \Delta_\gamma^\alpha=-\sum_b \mathfrak{D}_{ab} \left(q_a^\alpha+\delta q_{ab}^\alpha \right),
\end{eqnarray}
where $A^{\left<\alpha\beta\right>}$ is the traceless tensor and $D\equiv U_a^\alpha \partial_\alpha$. Of course, terms $\delta \Pi_{aa}=\delta \Pi^{<\alpha\beta>}_{aa}=\delta q^\alpha_{aa}=0$ in Eqs. (19)-(21), because diffusive effects never emerge in binary collision between two hard spherical particles, which belong to same species.\\
Multiplying $\Delta^\alpha_\beta$ or ${U_a}_\beta$ by both sides of Eq. (4) with Eq. (17), we obtain following relations by neglecting nonlinear terms
\begin{eqnarray}
&&\frac{n_a h_a}{c^2} DU_a^\alpha=\nabla^\alpha\left(p_a+\Pi_a\right)-\nabla_\beta \Pi_a^{\left<\alpha\beta\right>}-\frac{1}{c^2} D q_a^\alpha+\Delta^\alpha_\beta {\Psi_a}^\beta, \\
&&n_a D e_a=-p_a \nabla_\beta U_a^\beta-\nabla_\beta q_a^\beta+{U_a}_\beta {\Psi_a}^\beta,
\end{eqnarray}
where $h_a=mc^2 G_a$ ($G_a\equiv K\left(\chi_a\right)/K_2\left(\chi_a\right)$) is the enthalpy density.\\
From Eq. (3), we readily obtain following relation using Eq. (16)
\begin{eqnarray}
D n_a+n_a \nabla^\alpha {U_a}_\alpha=0.
\end{eqnarray}
Substituting Eqs. (22)-(24) into Eqs. (19) and (21), we can rewrite Eqs. (19) and (21) as
\begin{eqnarray}
&&\frac{C_2}{2} D \Pi_a-\frac{1}{2}n_a\left(m^2+C_1\right) \nabla^\alpha {U_a}_\alpha+\frac{\chi_a}{2 k \theta_a} \frac{C_1^\prime}{\chi_a^2+5\chi_a G_a-\chi_a^2 G_a^2-1} \nonumber \\
&&\times \left(\nabla^\alpha {q_a}_\alpha+p_a \nabla^\alpha {U_a}_\alpha-{U_a}_\alpha \Psi_a^\alpha\right)\nonumber \\
&&-5 \frac{C_3}{c^2} \nabla^\alpha {q_a}_\alpha+\frac{1}{6}\left(n_a m^2+5 n_a C_1\right)\nabla^\alpha {U_a}_\alpha=-\frac{3}{c^2} \sum_b \mathfrak{B}_{ab} \left(\Pi_a+\delta \Pi_{ab} \right),\\
&&5 C_3 D q_a^\alpha-\frac{c^4}{6}\left[\left(m^2-C_1\right)\nabla^\alpha n_a+\frac{\chi_a}{\theta_a} n_a C_1^\prime \nabla^\alpha \theta_a-C_2 \nabla^\alpha \Pi \right]-c^2 C_4 \nabla_\beta \Pi_a^{\left<\alpha \beta\right>} \nonumber\\
&&-\frac{c^4}{6 n_a h_a} \left(n_a m^2+5 n_a C_1\right) \left[\nabla^\alpha\left(p_a+\Pi_a\right)-\nabla_\beta \Pi_a^{\left<\alpha\beta\right>}-\frac{1}{c^2} D q_a^\alpha+\Delta^\alpha_\beta {\Psi_a}^\beta\right] \nonumber \\
&&=-\sum_b \mathfrak{D}_{ab} \left(q_a^\alpha+\delta q_{ab}^\alpha \right).
\end{eqnarray}
The first Maxwellian iteration of Eqs. (20), (25) and (26) \textcolor{black}{yields} the NSF law by setting $\Pi_a=\Pi_a^{\left<\alpha\beta\right>}=q_a^\alpha=0$ in left hand sides of Eqs. (20), (25) and (26) and $\Pi_a=\left[\Pi_a\right]_{\mbox{\tiny{NSF}}}$, $\Pi_a^{\left<\alpha\beta\right>}=\left[\Pi_a^{\left<\alpha\beta\right>}\right]_{\mbox{\tiny{NSF}}}$ and $q_a^\alpha=\left[q_a^\alpha\right]_{\mbox{\tiny{NSF}}}$ in right hand sides of Eqs. (20), (25) and (26), when we assume that $\left|\left[\Pi_b\right]_{\mbox{\tiny{NSF}}}-\left[\Pi_a\right]_{\mbox{\tiny{NSF}}}\right| \ll \left|\left[\Pi_a\right]_{\mbox{\tiny{NSF}}}\right|$, $\left|\left[\Pi_b^{\left<\alpha\beta\right>}\right]_{\mbox{\tiny{NSF}}}-\left[\Pi_a^{\left<\alpha\beta\right>}\right]_{\mbox{\tiny{NSF}}}\right| \ll \left|\left[\Pi_a^{\left<\alpha\beta\right>}\right]_{\mbox{\tiny{NSF}}}\right|$, and $\left|\left[q_b^{\alpha}\right]_{\mbox{\tiny{NSF}}}-\left[q_a^{\alpha}\right]_{\mbox{\tiny{NSF}}}\right| \ll \left|\left[q_a^\alpha\right]_{\mbox{\tiny{NSF}}}\right|$, as discussed in appendix A, as follows
\begin{eqnarray}
\left[\Pi_a\right]_{\mbox{\tiny{NSF}}}&=&-\eta_{a} \nabla_\alpha {U_a}^\alpha+\mathscr{L}_a,\\
\left[\Pi_a^{\left<\alpha\beta\right>}\right]_{\mbox{\tiny{NSF}}}&=& 2 \mu_{a} \nabla^{<\alpha} U_a^{\beta>}+\mathscr{M}_a^{\left<\alpha\beta\right>}\\
\left[q_a^\alpha \right]_{\mbox{\tiny{NSF}}}&=&\lambda_{a}\left(\nabla^\alpha \theta_a-\frac{\theta_a}{n_ah_a} \nabla^\alpha p_a\right)+\mathscr{N}_a^\alpha,
\end{eqnarray}
where $\eta_{a}$ (bulk viscosity), $\mu_{a}$ (viscosity coefficient) and $\lambda_{a}$ (thermal conductivity) are written as
\begin{eqnarray}
\eta_{a}&=&\frac{c^2}{3}\frac{n_a m^2}{\sum_b \mathfrak{B}_{ab}}\frac{20 G_a+2\chi_a^2 G_a^3+3\chi_a-13\chi_a G_a-2\chi_a^2 G_a}{\chi_a^3 G_a^2+\chi_a-\chi_a^3-5\chi_a^2 G_a},\\
\mu_{a}&=& \frac{n_a m^2 c^2}{\sum_b \mathfrak{C}_{ab}}\frac{G_a}{\chi_a},\\
\lambda_{a}&=&\frac{n_a m^2 c^4}{\sum_b \mathfrak{D}_{ab}}\frac{G_a^\prime}{\theta_a},
\end{eqnarray}
where $G^\prime_a=dG_a/d\chi_a$.\\
Additionally, $\mathscr{L}_a$, $\mathscr{M}_a^{\left<\alpha\beta\right>}$ and $\mathscr{N}_a^\alpha$ in Eqs. (27)-(29) are written as
\begin{eqnarray}
\mathscr{L}_a&=&\frac{1}{\sum_b \mathfrak{B}_{ab}} \left(\sum_b \mathfrak{B}_{ab} \delta \Pi_{ab}+\frac{1}{6 m} \frac{\chi_a^2 C_1^\prime}{\chi_a^2+5\chi_a G_a-\chi_a G_a^2-1}{U_a}_\alpha \left[\Psi_a^\alpha\right]^{(0)}\right),\\
\mathscr{M}_a^{\left<\alpha\beta\right>}&=&\frac{\sum_b \mathfrak{C}_{ab} \delta \Pi_{ab}^{\left<\alpha\beta\right>}}{\sum_b \mathfrak{C}_{ab}},\\
\mathscr{N}_a^\alpha &=& \frac{1}{\sum_b \mathfrak{D}_{ab}} \left[\sum_b \mathfrak{D}_{ab} \delta \Pi_{ab}+m c^2\left(1+\frac{5 G_a}{\chi_a}\right)\Delta_\beta^\alpha \left[\Psi_a^\beta\right]^{(0)}\right],
\end{eqnarray}
where $\left[\Psi_a^\alpha\right]^{(0)}$ is obtained by setting $\left[\Pi_i\right]_{\mbox{\tiny{NSF}}}=\left[\Pi_i^{\left<\alpha\beta\right>}\right]_{\mbox{\tiny{NSF}}}=\left[q_i^\alpha\right]_{\mbox{\tiny{NSF}}}=0$ ($i=a,b$) in $\Psi_a^\alpha$ \cite{memo3}, and $\mathscr{L}_a$, $\mathscr{M}_a^{\left<\alpha\beta\right>}$ and $\mathscr{N}_a^\alpha$ are finite terms in the NSF law in Eqs. (27)-(29), which are derived from the diffusion between species \enquote{$a$} and \enquote{$b$}. In the nonrelativistic gas, concrete forms of $\mathscr{L}_a$, $\mathscr{M}_a^{\left<\alpha\beta\right>}$ and $\mathscr{N}_a^\alpha$ were calculated for the inelastic Maxwell model using \textcolor{black}{five field variables}, namely, $n_i$, $\bm{u}_i$ and $\theta_i$ by Garzo and Astillero \cite{Garzo} (i.e., see Eq. (B22) in Ref. \cite{Garzo} for $\mathscr{N}_a^\alpha$), whereas the NSF law for the relativistic binary mixture gas is obtained using Israel-Stewart particles, when $\bm{u}_a=\bm{u}_b=\bar{\bm{u}}$, $\theta_a=\theta_b=\bar{\theta}$, and $m_a \simeq m_b$ \cite{Kremer}. Provided that $f_a=f_b$ and $d_a=d_b$, transport coefficients in Eqs. (30)-(32) coincide with those for the single component gas, and the NSF law for the single component gas is reproduced, because $\mathscr{L}_a=\mathscr{M}_a^{\left<\alpha\beta\right>}=\mathscr{N}_a^\alpha=0$ is obtained in Eqs. (27)-(29). Concrete forms of $\mathscr{L}_a$, $\mathscr{M}_a^{\left<\alpha\beta\right>}$ and $\mathscr{N}^\alpha$ in Eqs. (33)-(35) are obtained, when we calculate $\delta \Pi_a$, $\delta \Pi^{\left<\alpha\beta\right>}_a$, $\delta q_a^\alpha$ and $\Psi_a^\alpha$ in Eqs. (33)-(35). Such calculations, however, involve mathematical difficulties in our present study. Similarly, calculations of $\mathfrak{B}_{ab}$, $\mathfrak{C}_{ab}$ and $\mathfrak{D}_{ab}$ also involve mathematical difficulties. Then, we make two assumptions to compare later numerical results of $\Pi_a$, $\Pi_a^{\left<\alpha\beta\right>}$ and $q_a^\alpha$ with their analytical results. One is $\left|\mathscr{L}_a\right| \ll \left|-\eta_{a} \nabla_\alpha {U_a}^\alpha\right|$, $\left|\mathscr{M}_a^{\left<\alpha\beta\right>}\right| \ll \left| 2 \mu_{a} \nabla^{<\alpha} U_a^{\beta>}\right|$ and $\left|\mathscr{N}^\alpha\right| \ll \left| \lambda_{a}\left(\nabla^\alpha \theta_a-\theta_a/\left(n_ah_a\right)\nabla^\alpha p_a\right)\right|$ in Eqs. (27)-(29).\\
In other words, Eqs. (27)-(29) are reduced to
\begin{eqnarray}
&&\left[\Pi_a\right]_{\mbox{\tiny{NSF}}} \simeq-\eta_{a} \nabla_\alpha {U_a}^\alpha,\\
&&\left[\Pi_a^{\left<\alpha\beta\right>}\right]_{\mbox{\tiny{NSF}}} \simeq 2 \mu_{a} \nabla^{<\alpha} U_a^{\beta>},\\
&&\left[q_a^\alpha \right]_{\mbox{\tiny{NSF}}} \simeq \lambda_{a}\left(\nabla^\alpha \theta_a-\frac{\theta_a}{n_a h_a} \nabla^\alpha p_a\right).
\end{eqnarray}
 The other is that $\mathfrak{B}_{ab}$, $\mathfrak{C}_{ab}$ and $\mathfrak{D}_{ab}$ in Eqs. (30)-(32) are obtained by replacing $n_a$, $\chi_a$ and $d_a$ in $\mathfrak{B}_{aa}$, $\mathfrak{C}_{aa}$ and $\mathfrak{D}_{aa}$ \cite{Cercignani} with $n_b$, $\chi_{ab}$ ($\theta_{ab}$) and $d_{ab}$ as follows:
\begin{eqnarray}
\mathfrak{B}_{ab}&=&\frac{64\pi n_b k \theta_{ab}\sigma_{ab}}{3 c K_2\left(\chi_{ab}\right)^2\chi_{ab}^3}\frac{\left(1-\chi_{ab}^2-5\chi_{ab} G_{ab}+\chi_{ab}^2 G_{ab}^2\right)\left[2 K_2\left(2\chi_{ab}\right)+\chi_{ab}K_3\left(2\chi_{ab}\right)\right]}{20 G_{ab}+3\chi_{ab}-13 G_{ab}^2 \chi_{ab}-2 G_{ab}\chi_{ab}^2+2 G_{ab}^3\chi_{ab}^2}, \nonumber \\
\mathfrak{C}_{ab}&=&\frac{64\pi n_b k \theta_{ab}\sigma_{ab}}{15 c K_2\left(\chi_{ab}\right)^2\chi_{ab}^3 G_{ab}}\left[\left(2+\chi_{ab}^2\right)K_2\left(2\chi_{ab}\right)+\left(3\chi_{ab}^3+49\chi_{ab}\right)K_3\left(2\chi_{ab}\right)\right],\nonumber \\
\mathfrak{D}_{ab}&=&-\frac{64\pi n_b k \theta_{ab}\sigma_{ab}}{3 c K_2\left(\chi_{ab}\right)^2\chi_{ab}^3}\frac{\left(2+\chi_{ab}^2\right)K_2\left(2\chi_{ab}\right)+5\chi_{ab}K_3\left(2\chi_{ab}\right)}{\chi_{ab}+5 G_{ab}-\chi_{ab} G_{ab}^2},
\end{eqnarray}
where $G_{ab}=K_3\left(\chi_{ab}\right)/K_2\left(\chi_{ab}\right)$ and $\sigma_{ab}=d_{ab}^2/4$. We mention to $\chi_{ab}(=k \theta_{ab}/(mc^2))$ in Sec. III.\\
Finally, we remind that Eqs. (36)-(38) correspond to markedly simplified NSF law, because we neglected all the diffusive terms in the NSF law. As mentioned in appendix A, the NSF law of the species \enquote{$a$} depends on gradients of five field variables of other species. Meanwhile, we \textcolor{black}{can confirm effects via diffusive terms in the NSF law, when the NSF law in Eqs. (36)-(38) for the binary mixture gas with equal masses approximates dissipating terms with worse accuracies than the NSF law for the single component gas does, because such a worse approximation by the NSF law in Eqs. (36)-(38) is caused by neglect of the diffusive terms in the NSF law in Eqs. (36)-(38).} Actually, later numerical result confirms that the Fourier law in Eq. (38) for the binary mixture gas with equal masses approximates \textcolor{black}{$q_a^\alpha$ ($a=A,B$)} with worse accuracies than the NSF law for the single component gas does.
\subsection{NSF order approximation of diffusion flux for binary mixture gas with equal masses \cite{Kox}}
We mention to \textcolor{black}{the} diffusion and thermal diffusion coefficients for the binary mixture gas with equal masses, which were calculated by Kox \textit{et al}. \cite{Kox}. Therefore, the author recommends readers to follow the mathematical procedures to calculate the diffusion and thermal diffusion coefficients in their original paper \cite{Kox}.\\
Setting $m_a=m_b=m$ in the binary mixture gas (species \enquote{$a$} and \enquote{$b$}), the diffusion and thermal-diffusion coefficients were calculated by Kox \textit{et al}. \cite{Kox} using the approximation on the basis of Laguerre polynomials as
\begin{eqnarray}
\left[\mathcal{D}\right]_1&=&\frac{3 c}{n \pi d_{ab}^2}\frac{K_2(\chi)^2}{(8+4\chi^{-2})K_2(2\chi)+28 \chi^{-1} K_3(2\chi)}, \nonumber \\
\left[\mathcal{D}\right]_2&=&\left[\mathcal{D}\right]_1+\frac{3}{8\pi}\frac{c}{n d_{ab}^2} \frac{\left(Q+2R\right)\left[T_1\left(\chi\right)K_2\left(\chi\right)-T_2\left(\chi\right)K_2\left(\chi\right)\right]^2}{T_2\left(\chi\right)\left[P N_1\left(\chi\right)+Q N_1\left(\chi\right)+R N_3\left(\chi\right) \right]},\nonumber \\
\left[\mathcal{D}_T\right]_1&=&\frac{3}{8\pi} \frac{k}{c} \frac{c_p}{c_p-c_v} \nonumber \\
&&\times \frac{\left[\left(\rho_a-\rho_b\right)d_{ab}^2+\rho_a d_{a}^2-\rho_{b} d_{b}^2\right]K_2\left(\chi\right)\left[T_1\left(\chi\right)K_2\left(\chi\right)-T_2\left(\chi\right)K_2\left(\chi\right)\right]}{P N_1\left(\chi\right)+Q N_1\left(\chi\right)+R N_3\left(\chi\right)},
\end{eqnarray}
where $d_{ab}=\left(d_a+d_b\right)/2$, $P \equiv \rho_a \rho_b d_{a}^2 d_{b}^2$, $Q \equiv (\rho_a^2 d_{a}^2+\rho_{b}^2 d_{b}^2)d_{ab}^2$ and $R\equiv \rho_a \rho_b d_{ab}^4$, in which $\rho_a=m n_a$ ($\rho_b=m n_b$) is the density of the species $a$ ($b$), and $d_a$ ($d_b$) is the diameter of the species $a$ ($b$). Subscriptions $n$ in $\left[\mathcal{D}\right]_n$ and $\left[\mathcal{D}_T\right]_n$ ($n \in \mathbb{N}$) reveal orders of approximations with Laguerre polynomials. Here, we must remind that Kox \textit{et al}. \textcolor{black}{considered} the case of $U_a^\alpha=U_b^\alpha=\bar{U}^\alpha$ and $\theta_a=\theta_b=\bar{\theta}$ \cite{Kox}. In Eq. (40), $c_p$ ($c_v$) is the heat capacity per a hard spherical particle at the constant pressure (constant volume) \cite{Cercignani}. In Eq. (40), $T_1\left(\chi\right)$, $T_2\left(\chi\right)$, $N_1\left(\chi\right)$, $N_2\left(\chi\right)$ and $N_3\left(\chi\right)$ are defined as \cite{Kox}
\begin{eqnarray}
T_1\left(\chi\right)&=&\left(10 \chi^{-1}+4\chi^{-3}\right) K_2\left(2\chi\right)+\left(2+34\chi^{-2}\right) K_3\left(2\chi\right), \nonumber \\
T_2\left(\chi\right)&=&\left(2 + \chi^{-1}\right)K_2\left(2\chi\right)+7\chi^{-1}K_3\left(2\chi\right),\nonumber\\
N_1\left(\chi\right)&=&\left(4\chi^{-2}+10\chi^{-4}+4\chi^{-6}\right)K_2^2\left(2\chi\right)
+\left(34 \chi^{-3}+38\chi^{-5}\right) K_2 \left(2\chi\right) K_3 \left(2\chi\right) \nonumber \\
&&+ 70 \chi^{-4}K_3^2\left(2\chi\right),\nonumber\\
N_2\left(\chi\right)&=&\left(2 +11\chi^{-1}+14\chi^{-4}+4\chi^{-6}\right)K_2^2(2\chi)
+\left(5\chi^{-1}+75\chi^{-3}+50\chi^{-5}\right)K_2 \left(2\chi\right) K_3 \left(2\chi\right)\nonumber \\
&&+\left(-2-5\chi^{-2}+136\chi^{-4}\right)K_3^2\left(2\chi\right),\nonumber\\
N_3\left(\chi\right)&=&\left(4 +18\chi^{-2}+18\chi^{-4}+4\chi^{-6}\right)K_2^2\left(2\chi\right)
+\left(10 \chi^{-1}+116\chi^{-3}+ 62\chi^{-5}\right)K_2 \left(2\chi\right) K_3 \left(2\chi\right) \nonumber \\
&&+\left(-4-10\chi^{-2}+ 202\chi^{-4}\right) K_3^2\left(2\chi\right),
\end{eqnarray}
where we assume that $\chi$ is calculated by the energy conservation, namely, $n_a e_a+n_b e_b=\left(n_a+n_b\right)e$, even when $U_a^\alpha \neq U_b^\alpha$, where $e=mc^2\left(G-1/\chi\right)$ ($G\equiv K_3\left(\chi\right)/K_2\left(\chi\right)$) is the averaged energy density.\\
Finally, the diffusion flux ($J_{ab}^\alpha$) between species \enquote{$a$} and \enquote{$b$} is obtained as
\begin{eqnarray}
J_{ab}^\alpha=-J_{ba}^\alpha=-\rho c_a c_b \mathcal{D}_T \bar{\nabla}^\alpha \bar{\theta}-\rho \mathcal{D} \bar{\nabla}^{\alpha} c_a,
\end{eqnarray}
where $\bar{\nabla}^\alpha=\bar{\Delta}^{\alpha\beta} \partial_\beta$, $\rho=\rho_a+\rho_b$ and $c_a=n_a/\left(n_a+n_b\right)$, and $\bar{\theta}$ is the averaged temperature over species \enquote{$a$} and \enquote{$b$}.\\
From the definition of the diffusion flux $J_a^\alpha$ in Eq. (13), we can readily understand that \textcolor{black}{$J_a^\alpha$ is not a dissipating term, which depends on the generic Knudsen number, because $N_a^\alpha$ in Eq. (10) does not include dissipating terms, which depend on the generic Knudsen number}. As a result, $J_a^\alpha$ depends on the difference between $\bar{u}^i$ \textcolor{black}{($\bar{U}^\alpha$)} and $u_a^i$ \textcolor{black}{($U_a^\alpha$)}, exclusively. In previous studies on the diffusion flux \cite{Kremer} \cite{Kox}, $J_a^\alpha$ is approximated with gradients of the number density fraction ($c_a$) and temperature, as shown in Eq. (42), because \textcolor{black}{the diffusion flux was regarded as a dissipating term, which depends on the generic Knudsen number \cite{Yano2}}. In later discussions on numerical results, we certainly confirm that $J_a^\alpha$ is not sensitive to the increase of the generic Knudsen number \textcolor{black}{unlike other dissipating terms such as $\Pi_a$, $\Pi_a^{\left<\alpha\beta\right>}$ and $q_a^\alpha$. We, however, remind that the diffusion flux is surely dissipated by binary collisions between species \enquote{$a$} and \enquote{$b$}, because the momentum transfer between two different species of particles via binary collisions decreases the difference between $U_a^\alpha$ and $U_b^\alpha$, (namely, $U_a^\alpha, U_b^\alpha \rightarrow \bar{U}^\alpha$ via binary collisions between species \enquote{$a$} and \enquote{$b$}).}
\section{Numerical analysis of rarefied shock layer of binary mixture gas}
In this section, we investigate the characteristics of dissipation process of the thermally relativistic flow of the binary mixture gas, which is composed of two species of hard spherical particles with equal masses and different diameters, numerically. As an object of the numerical analysis, we investigate the rarefied shock layer around the triangle prism, as shown in Fig. 1, because such a problem has been discussed in our previous studies \cite{Yano1} \cite{Yano2} of the thermally relativistic flow of the single component gas. The vertical angle of the triangle prism is set as $120$ degrees, whereas the upper-half of the triangle prism ($0 \le Y$) is analyzed owing to symmetries of numerical results at both sides of $Y=0$, as shown in Fig. 1. In our later discussion, the quantity with a bracket $[]$ indicates the approximated value obtained using the analytical result in Sec. II, whereas the quantity without a bracket indicates the numerical value, which is obtained by solving the RBE.
\begin{center}
\includegraphics[width=0.75\textwidth]{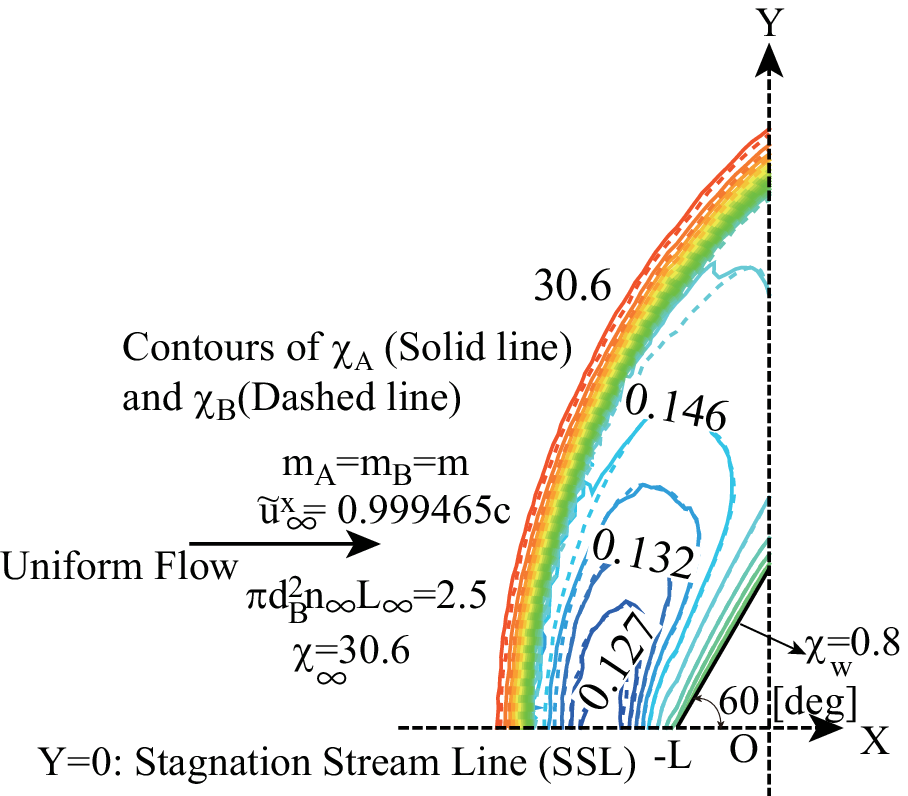}\\
\small{FIG. 1: Schematic of flow-field in Test (II).}
\end{center}
\subsection{Dissipation process of binary mixture gas with equal masses}
Firstly, we investigate dissipation process of the thermally relativistic flow of the binary mixture gas (species A and B), when masses of two species, A and B, are equal, namely, $m_A=m_B$. Then, we consider two types of uniform flows in Tests (I) and (II). In Test (I), the uniform flow corresponds to the mildly thermally relativistic flow with the small Lorentz contraction, in which the temperature of the uniform flow is set as ${\chi_A}_\infty={\chi_B}_\infty=\chi_{\infty}=47$, and the flow velocity of the uniform flow is set as ${u_A^x}_\infty={u_B^x}_\infty=u^x_\infty=0.6 c$, where symbols with the subscription $\infty$ indicates quantities in the uniform flow. In Test (II), the uniform flow corresponds to the mildly thermally relativistic flow with the large Lorentz contraction, in which the temperature of the uniform flow is set as ${\chi_A}_\infty={\chi_B}_\infty=\chi_{\infty}=30.6$, and the velocity of the uniform flow is set as ${u_A^x}_\infty={u_B^x}_\infty=u^x_\infty=0.999465c$. The schematic of the flow-field in Test (II) is shown in Fig. 1. In both Tests (I) and (II), $d_A/d_\infty=0.5$, $d_B/d_\infty=1$, $m_A=m_B=m$ and ${n_A}_\infty={n_B}_\infty=n_\infty$, whereas the scale parameter, $n_\infty \pi d_B^2 L_\infty$, is set as $2.5$. In later descriptions, symbols with $\tilde{}$ indicate nondimensionalized quantities, such as $\tilde{m}_i=m_i/m$, $\tilde{d}_i=d_i/d$ ($i$=A,B), $\tilde{x}^i=x^i/L$, $\tilde{v}^i=v^i/c$, $\tilde{u}^i=u^i/c$, $\tilde{\theta}=\theta/\theta_\infty$, $\tilde{J}^{\alpha}=J^\alpha/\left(n_\infty c\right)$, $\tilde{\Pi}=\Pi/\left(n_\infty m c^2\right)$, $\tilde{\Pi}^{\left<\alpha\beta\right>}=\Pi^{\left<\alpha\beta\right>}/\left(n_\infty mc^2\right)$ and $\tilde{q}^\alpha=q^\alpha/\left(n_\infty mc^3\right)$. As shown in Fig. 1, we consider the Cartesian coordinate $\left(x^1,x^2\right) \rightarrow \left(X,Y\right)$ in the laboratory frame. In particular, physical quantities along the stagnation stream line (SSL) ($Y=0$ $\wedge$ $X \le -L$) are discussed. As a numerical method to solve the RBE for the binary mixture gas, the DSMC method \cite{Bird} is used. $\left(X,Y\right)=\left(48,80\right)$ girds are equally spaced in $-4L \le X \le -L$, and $0 \le Y \le 8L$ and $\left(X,Y\right)=(12,80)$ girds are equally spaced in $-L \le X \le 0$ and $-\sqrt{3}/{2}X \le Y \le 8L-\sqrt{3}/{2}X$. In both Tests (I) and (II), about 130 sample particles are set per unit cell in the uniform flow. The hard spherical particles, which collide with the wall, are reflected from the wall with the thermally equilibrium state, which is defined by Maxwell-J$\ddot{\mbox{u}}$ttner function, whose flow velocity and temperature are zero and $\theta_w$ (wall temperature), respectively. In Test (I), the wall temperature is set as $\chi_w=mc^2/\left(k\theta_w\right)=30$. In Test (II), the wall temperature is set as $\chi_w=0.8$.\\
Figure 2 shows profiles of the number density, flow velocity and temperature along the SSL in Test (I) (upper frame) and Test (II) (lower frame). The shock wave separation \cite{Rebrov} between two species is confirmed in profiles of the number density, velocity and temperature in both Tests (I) and (II), as shown in both frames of Fig. 2. Additionally, the thickness of the shock wave, which is obtained for the species A, is thicker than that for the species B in both Tests (I) and (II), because the mean free path of the species A is longer than that of the species B. Meanwhile, marked differences between $\tilde{n}_A$ and $\tilde{n}_B$, or $\tilde{u}_A^x$ and $\tilde{u}_A^x$ are confirmed in the thermal boundary layer behind the shock wave, namely, $-X/L \le 1.7$ in Test (II). As a result, species A and B are under the nonequilibrium state in the thermal boundary layer in Test (II), whereas we conjecture that species A and B are similar to the equilibrium state in the thermal boundary layer ($-X/L \le 2.5$) in Test (I). Thus, the increase of the Lorentz contraction in the uniform flow yields the increase of the nonequilibrium in the thermal boundary layer. Such an increase of the nonequilibrium in the thermal boundary layer in accordance with the increase of the Lorentz contraction in the uniform flow was surely confirmed in our previous study of the single component gas \cite{Yano2}, whereas the present study indicates that the nonequilibrium between two species also increases via the increase of the Lorentz contraction in the uniform flow \cite{memo}.
\begin{center}
\includegraphics[width=0.75\textwidth]{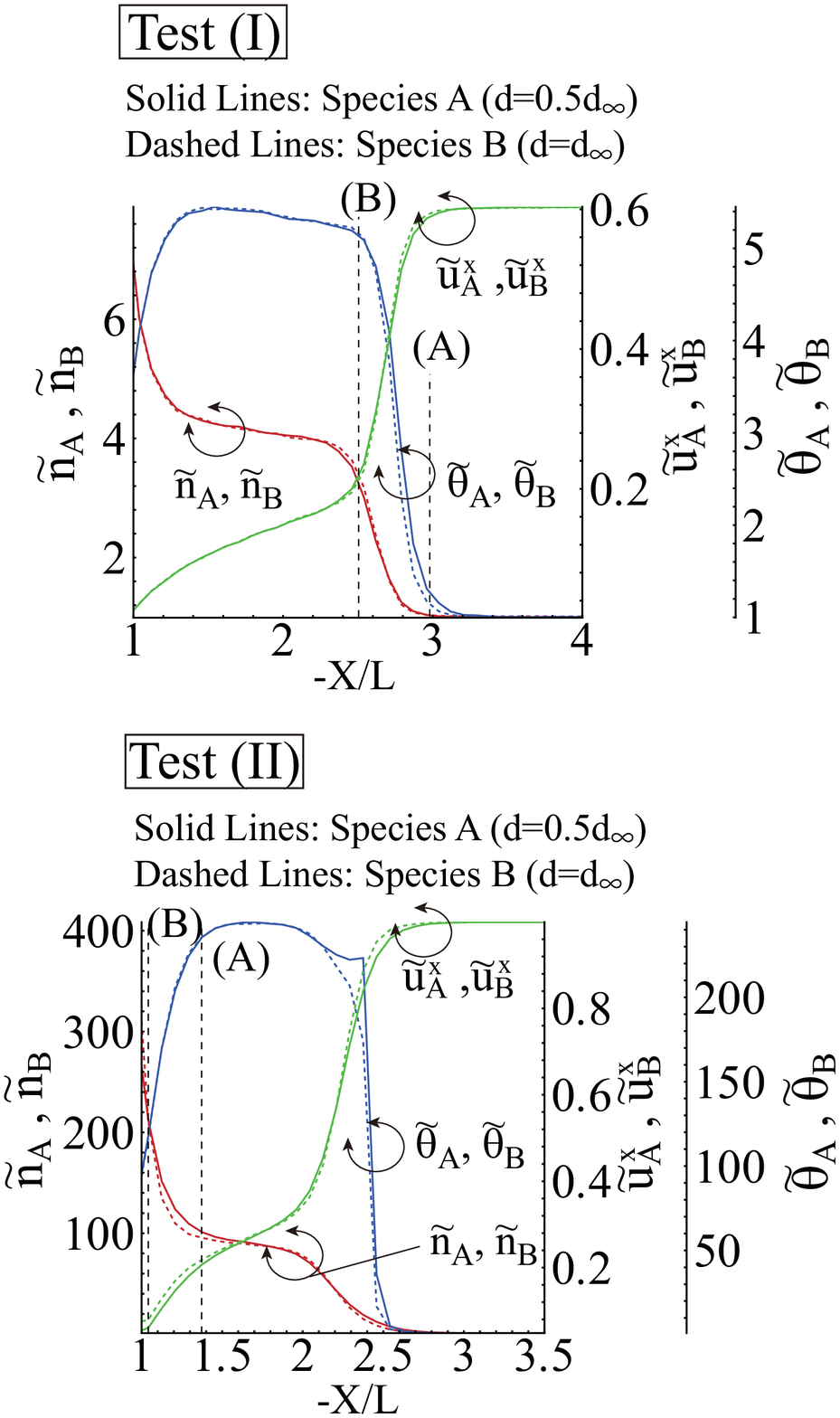}\\
\small{FIG. 2: Profiles of $\tilde{n}_{i}$, $\tilde{u}^x_i$ and $\tilde{\theta}_i$ ($i=$A, B) along the SSL in Tests (I) (upper frame) and (II) (lower frame).}
\end{center}
Figure 3 shows profiles of the diffusion flux along the SSL in Tests (I) (upper frame) and (II) (lower frame). Here, we define $\left[J_{AB}^\alpha\right]_1$, $\left[J_{AB}^\alpha \right]_2$ and $\left[J_{AB}^\alpha\right]_T$ to investigate the accuracy of the approximation with Laguerre polynomials in Eq. (42) and effects via the thermal-diffusion as
\begin{eqnarray}
\left[J_{AB}^\alpha\right]_1&=&-\rho c_A c_B \left[\mathcal{D}_T\right]_1 \bar{\nabla}^\alpha \bar{\theta}-\rho \left[\mathcal{D}\right]_1 \bar{\nabla}^{\alpha} c_A, \nonumber \\
\left[J_{AB}^\alpha\right]_2&=&-\rho c_A c_B \left[\mathcal{D}_T\right]_1 \bar{\nabla}^\alpha \bar{\theta}-\rho \left[\mathcal{D}\right]_2 \bar{\nabla}^{\alpha} c_A, \nonumber \\
\left[J_{AB}^\alpha\right]_T&=&-\rho c_A c_B \left[\mathcal{D}_T\right]_1 \bar{\nabla}^\alpha \bar{\theta},
\end{eqnarray}
where $\bar{\theta}$ is calculated using the relation $n_A e_A+n_B e_B=n e$, although the flow velocity of the species A is different from that of the species B.\\
In Test (I), $\left[\tilde{J}_{AB}^x\right]_T$ is dominant in $\left[\tilde{J}_{AB}^x\right]_1$ or $\left[\tilde{J}_{AB}^x\right]_2$ inside the shock wave ($2.5 \le -X/L \le 3.3$), whereas $\left[\tilde{J}_{AB}^x\right]_1 \simeq \left[\tilde{J}_{AB}^x\right]_2$ is obtained in all the domain. Additionally, $\left[\tilde{J}_{AB}^x\right]_T$ approximates to zero in the range of $1.4 \le -X/L \le 2.5$, whereas $0 \le \left[\tilde{J}_{AB}^x\right]_2$ is obtained behind the shock wave, namely, $2.2 \le -X/L \le 2.5$. The global tendency of the profile of $\tilde{J}_{AB}^x$ is similar to $\left[\tilde{J}_{AB}^x\right]_2$ in the range of $1.4 \le -X/L$, whereas the negative peak value of $\tilde{J}_{AB}^x$ is smaller than $\left[\tilde{J}_{AB}^x\right]_2$ inside the shock wave ($-X/L \simeq 2.6$) and positive peak of $\tilde{J}_{AB}^x$ is larger than $\left[\tilde{J}_{AB}^x\right]_2$ behind the shock wave ($-X/L \simeq 2.5$). Numerical results surely confirm the relation $\tilde{J}_{AB}^x=-\tilde{J}_{BA}^x$, as shown in both frames of Fig. 3. On the other hand, the positive signature of $\left[\tilde{J}_{AB}^x\right]_2$ is opposite to the negative signature of $J_{AB}^x$ in the thermal boundary layer ($-X/L \le 1.38$).\\
In Test (II), $\left[\tilde{J}_{AB}^x\right]_2$ has a positive peak around $-X/L=2.9$, which corresponds to the forward regime of the shock wave, as shown in the lower frame of Fig. 2. Meanwhile, $\left[\tilde{J}_{AB}^x\right]_T$ is dominant in $\left[\tilde{J}_{AB}^x\right]_2$ inside the shock wave ($2.3 \le -X/L \le 2.6$). $\left[\tilde{J}_{AB}^x\right]_2$ approximates to zero behind the shock wave ($1 \le -X/L \le 2.3$). $\tilde{J}_{AB}^x$ is approximately equal to zero in the forward regime of the shock wave ($3 \le -X/L$), whereas $\tilde{J}_{AB}^x$ has a negative peak around $-X/L\simeq 2.5$ and a positive peak around $-X/L \simeq 2.1$. $\tilde{J}_{AB}^x$ decreases toward the wall ($1 \le -X/L \le 2.1$). As a result, $\tilde{J}_{AB}^x$ never has a positive peak in the forward regime of the shock wave such as $\left[\tilde{J}_{AB}^x\right]_2$, whereas $\left[\tilde{J}_{AB}^x\right]_2$ never decreases toward the wall ($1 \le -X/L \le 2.1$) unlike $\tilde{J}_{AB}^x$.\\
In summary, the profile of the diffusion flux is roughly approximated using the NSF order approximation inside the shock wave in Test (I), as shown in the upper frame of Fig. 3, when the Lorentz contraction in the uniform flow is small. On the other hand, the magnitude of the diffusion flux, which is obtained using the DSMC method, is similar to that obtained using the NSF order approximation inside the shock wave in Test (II), when the Lorentz contraction in the uniform flow is large, as shown in the lower frame of Fig. 3. In other words, the magnitude of the diffusion flux is not sensitive to nonequilibrium terms beyond the NSF order approximation inside the shock wave, when the Lorenz contraction is large. The magnitude of the diffusion flux in the vicinity of the wall is, however, markedly larger than that obtained the NSF order approximation in Test (II), as shown in the lower frame of Fig. 3. In other words, the magnitude of the diffusion flux is sensitive to  nonequilibrium terms beyond the NSF order approximation in the vicinity of the wall, although effects via the Lorentz contraction are markedly small in the vicinity of the wall. Such a strong nonequilibrium in the vicinity of the wall is demonstrated by the fact that the nonequilibrium states between species A and B are not relaxed behind the shock wave owing to the thermally relativistic effects (see the lower frame of Fig. 6) together with nonequilibrium states owing to discontinuous distribution functions of species A and B on the wall \cite{Yano3}.
\begin{center}
\includegraphics[width=0.75\textwidth]{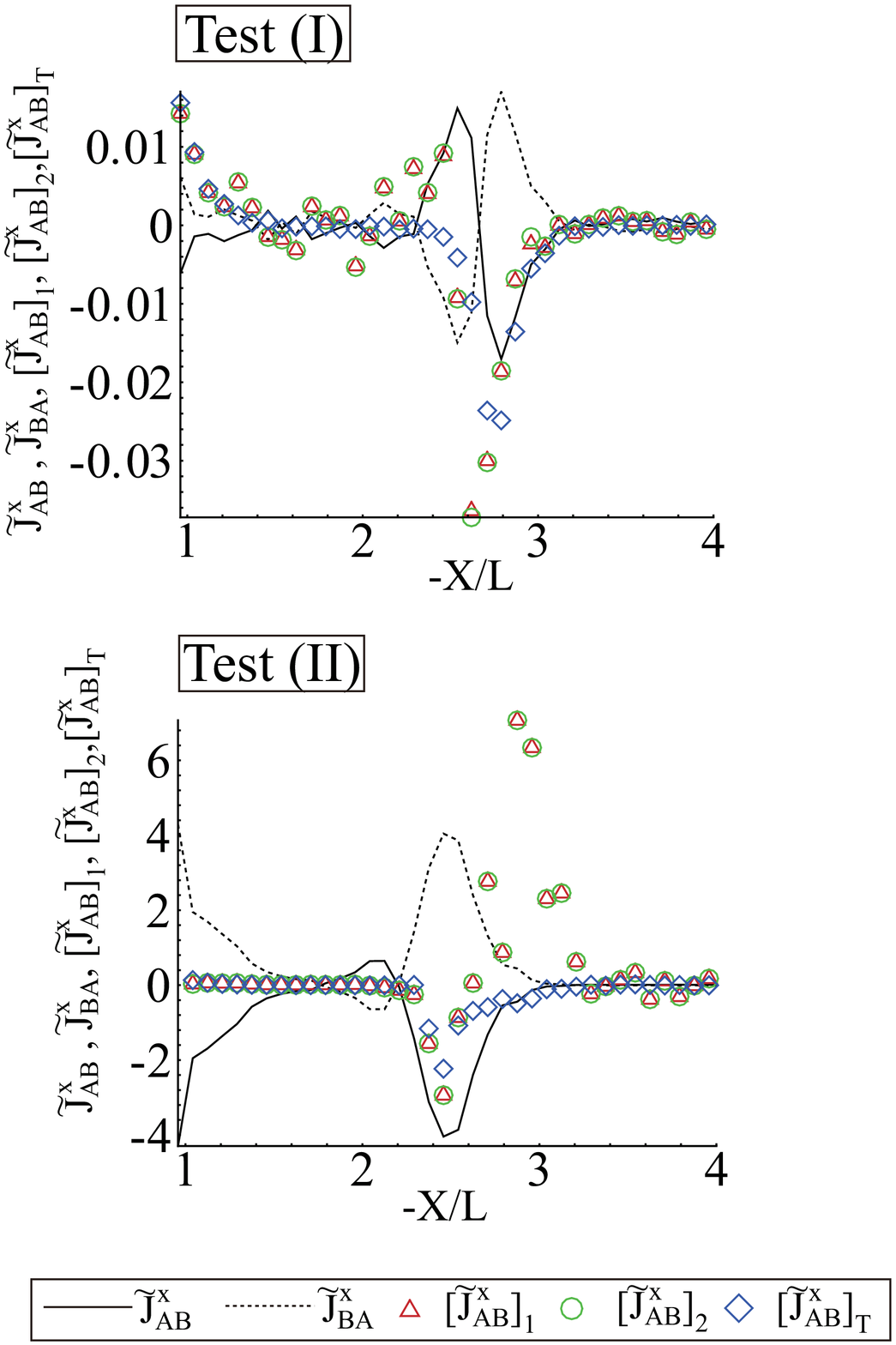}\\
\small{FIG. 3: Profiles of $\tilde{J}_{AB}^x$, $\tilde{J}_{BA}^x$, $\left[\tilde{J}_{AB}^x\right]_1$, $\left[\tilde{J}_{AB}^x\right]_2$ and $\left[\tilde{J}_{AB}^x\right]_T$ along the SSL in Tests (I) (upper frame) and (II) (lower frame).}
\end{center}
In above discussion on the diffusion flux, we postulated that the diffusion flux can be approximated using Chapman-Enskog approximation \cite{Shchavaliev} \cite{Zhdanov}, which also has been applied to the nonrelativistic mixture gas. On the other hand, we know that the diffusion flux does not depend on the generic Knudsen number \cite{Yano2} from Eq. (13). As discussed in Sec. II-(C), the diffusion flux depends on the difference between $\tilde{U}_A^x=\gamma(\tilde{u}_A^x)\tilde{u}_A^x$ and $\tilde{U}_B^x=\gamma(\tilde{u}_B^x)\tilde{u}_B^x$. Then, we define the approximate diffusive flux $\tilde{\mathfrak{J}}_{AB}^x$ as
\begin{eqnarray}
\tilde{\mathfrak{J}}_{AB}^x=\frac{1}{2}\tilde{\bar{n}} \left(\tilde{U}_A^x-\tilde{U}_B^x\right)=-\tilde{\mathfrak{J}}_{BA}^x,
\end{eqnarray}
where $\tilde{\bar{n}}=(\tilde{n}_A+\tilde{n}_B)/2$.\\
Figure 4 shows profiles of $\tilde{\mathfrak{J}}_{AB}^x$ along the SSL together with those of $\tilde{J}_{AB}^x$ in Tests I (upper frame) and II (lower frame). As shown in the upper frame of Fig. 4, $\tilde{\mathfrak{J}}_{AB}^x$ is markedly similar to $\tilde{J}_{AB}^x$. Such a similarity indicates that the diffusion flux can be demonstrated using $\tilde{\mathfrak{J}}_{AB}^x$ with a good accuracy. The upper frame of Fig. 4 shows that the difference between $\tilde{u}_A^x$ and $\tilde{u}_B^x$ contributes to $\tilde{J}_{AB}^x$ in the vicinity of the wall. The lower frame of Fig. 4 indicates that there are some differences between $\tilde{\mathfrak{J}}_{AB}^x$ and $\tilde{J}_{AB}^x$ in the forward regime of the shock wave ($2.44 \le -X/L \le 3.27$), whereas $\tilde{\mathfrak{J}}_{AB}^x$ is markedly similar to $\tilde{J}_{AB}^x$ in the range of $1 \le -X/L \le 2.44$. Therefore, $\tilde{\mathfrak{J}}_{AB}^x=\tilde{\bar{n}} \left(\tilde{U}_A^x-\tilde{U}_B^x\right)/2$ is a rough approximation of $\tilde{\mathfrak{J}}_{AB}^x$, when the local Lorentz contraction becomes large. As shown in the lower frame of Fig. 4, the difference between $\tilde{u}_A^x$ and $\tilde{u}_B^x$ contributes to $\tilde{J}_{AB}^x$ in the vicinity of the wall, exclusively. Finally, we have a mathematically open problem how the higher order approximation of $J_a^\alpha$ beyond the NSF approximation converges to $J_a^\alpha$, which is independent of the generic Knudsen number, when the generic Knudsen number increases in accordance with the increase of the Lorentz contraction in the uniform flow \cite{Yano2}.
\begin{center}
\includegraphics[width=0.75\textwidth]{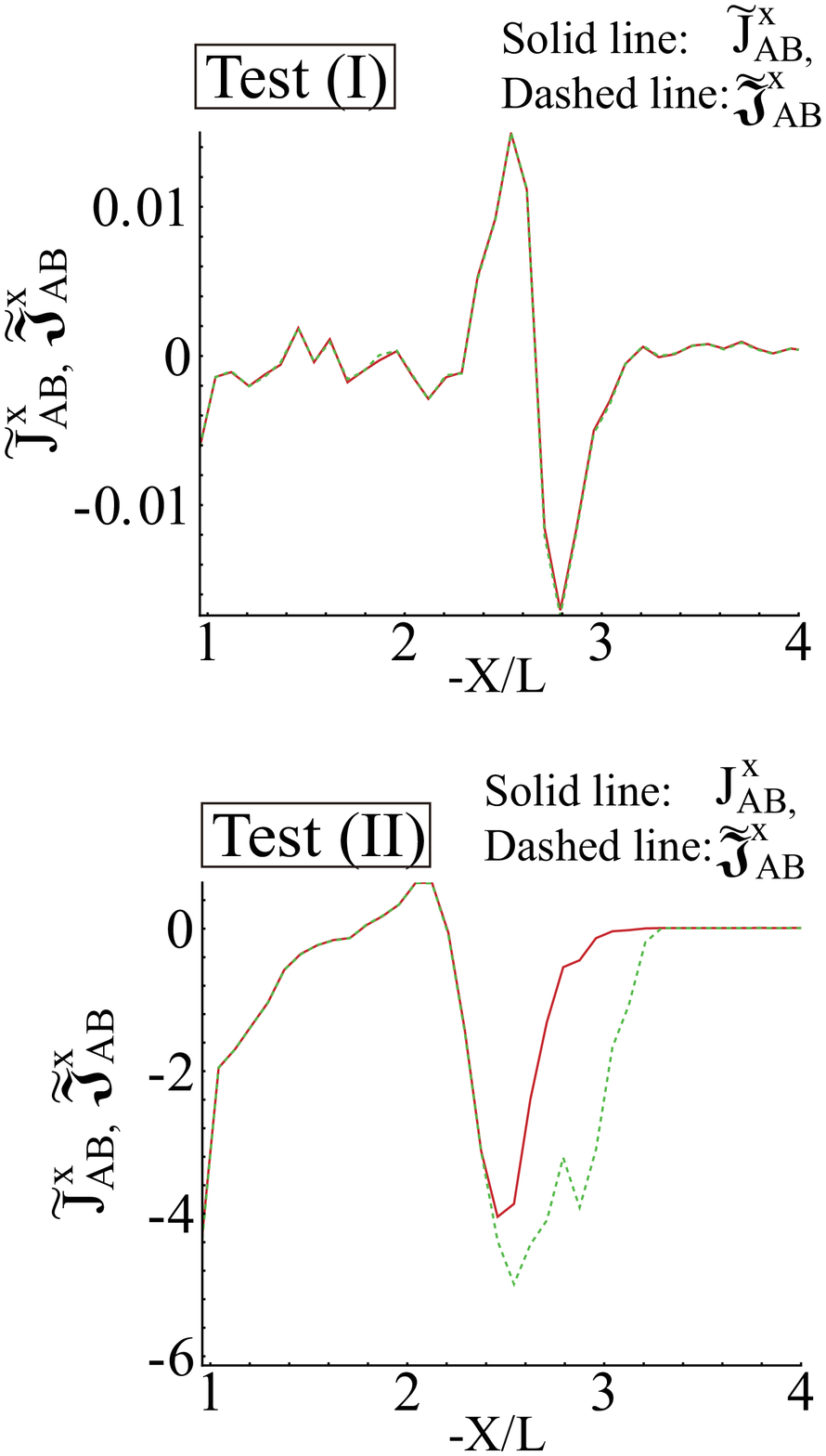}\\
\small{FIG. 4: Profiles of $\tilde{J}_{AB}^x$ and $\tilde{\mathfrak{J}}_{AB}^x$ along the SSL in Tests (I) (upper frame) and (II) (lower frame).}
\end{center}
\textcolor{black}{Figures 5-7} show profiles of $\tilde{\Pi}^{\left<xx\right>}_i$ and $\left[\tilde{\Pi}^{\left<xx\right>}_i\right]_{\mbox{\tiny{NSF}}}$, $\tilde{q}^x_i$ and $\left[\tilde{q}^{x}_i\right]_{\mbox{\tiny{NSF}}}$, and $\tilde{\Pi}_i$ and $\left[\tilde{\Pi}_i\right]_{\mbox{\tiny{NSF}}}$ along the SSL in Tests (I) (upper frames) and (II) (lower frames), respectively, where subscript $i$ corresponds to A or B. $\left[\tilde{\Pi}^{\left<xx\right>}_i\right]_{\mbox{\tiny{NSF}}}$, $\left[\tilde{\Pi}_i\right]_{\mbox{\tiny{NSF}}}$ and $\left[\tilde{q}^{x}_i\right]_{\mbox{\tiny{NSF}}}$ are calculated using Eqs. (30)-(32) and (36)-(39), in which we assume that $\chi_{AB}=\chi_{BA}=\chi=mc^2/\left(k\bar{\theta}\right)$. In Test (I), $\left[\Pi_i^{\left<xx\right>}\right]_{\mbox{\tiny{NSF}}} \le \Pi_i^{\left<xx\right>}$, $\left|\left[q_i^{x}\right]_{\mbox{\tiny{NSF}}}\right| \le \left|q_i^{x}\right|$ and $\left|\left[\Pi_i\right]_{\mbox{\tiny{NSF}}}\right| \le \left|\Pi_i\right|$ ($i=$A, B), as shown in upper frames of Figs. 5-7. In our previous study \cite{Yano2}, $\Pi$ for the single component gas can be approximated more accurately using its Burnett order approximation, whereas the calculation of Burnett order approximation of $\Pi_i$ is beyond our scope of this paper. Finally, we must refer to comparisons of profiles of heat fluxes in Test (I) with that for the single component gas, which \textcolor{black}{was} obtained using $\tilde{m}_A=\tilde{m}_B=1$ and $\chi_\infty=45$ (see the left frame of Fig. 2 in Ref. \cite{Yano2}). As shown in the left frame of Fig. 2 in Ref. \cite{Yano2}, profile of the heat flux for the single component gas is better approximated by the Fourier law in the thermal boundary layer than two heat fluxes in Test (I). \textcolor{black}{Consequently, we conclude that neglect of all the diffusive terms in the Fourier law in Eq. (29) degrades accuracies of approximations of two heat fluxes, whereas we must investigate whether the Navier-Stokes (NS) law for the pressure deviator in Eq. (37) approximates pressure deviators ($\Pi_{A}^{\left<xx\right>}$ and $\Pi_{B}^{\left<xx\right>}$) with worse accuracies than the NS law for the single component gas does, in our future study.} As shown in the upper frame of Fig. 3, $\tilde{J}_{AB}^x$ is nonzero value in the thermal boundary layer. Therefore, we conjecture that such worse NSF order approximations of two heat fluxes in the thermally boundary layer in Test (I) are caused by neglecting diffusive effects in the Fourier law in Eq. (29) (i.e., $\mathscr{N}_a^\alpha=0$ in Eq. (29) yields such worse NSF order approximations of heat fluxes). In Test (II), $\left[\Pi_i^{\left<xx\right>}\right]_{\mbox{\tiny{NSF}}} \ll \Pi_i^{\left<xx\right>}$, $\left|\left[q_i^{x}\right]_{\mbox{\tiny{NSF}}}\right| \ll \left|q_i^{x}\right|$ and $\left|\left[\Pi_i\right]_{\mbox{\tiny{NSF}}}\right| \ll \left|\Pi_i\right|$ ($i=$A, B) inside the shock wave and thermal boundary layer around the wall, as shown in lower frames of Figs. 5-7. Our previous studies \cite{Yano2} described that such marked differences are caused by the marked increase of the generic Knudsen number \cite{Yano2} owing to the marked increase of the Lorentz contraction. Consequently, terms beyond Burnett order terms are significant, when the flow velocity of the uniform flow approximates to the speed of light.
\begin{center}
\includegraphics[width=0.75\textwidth]{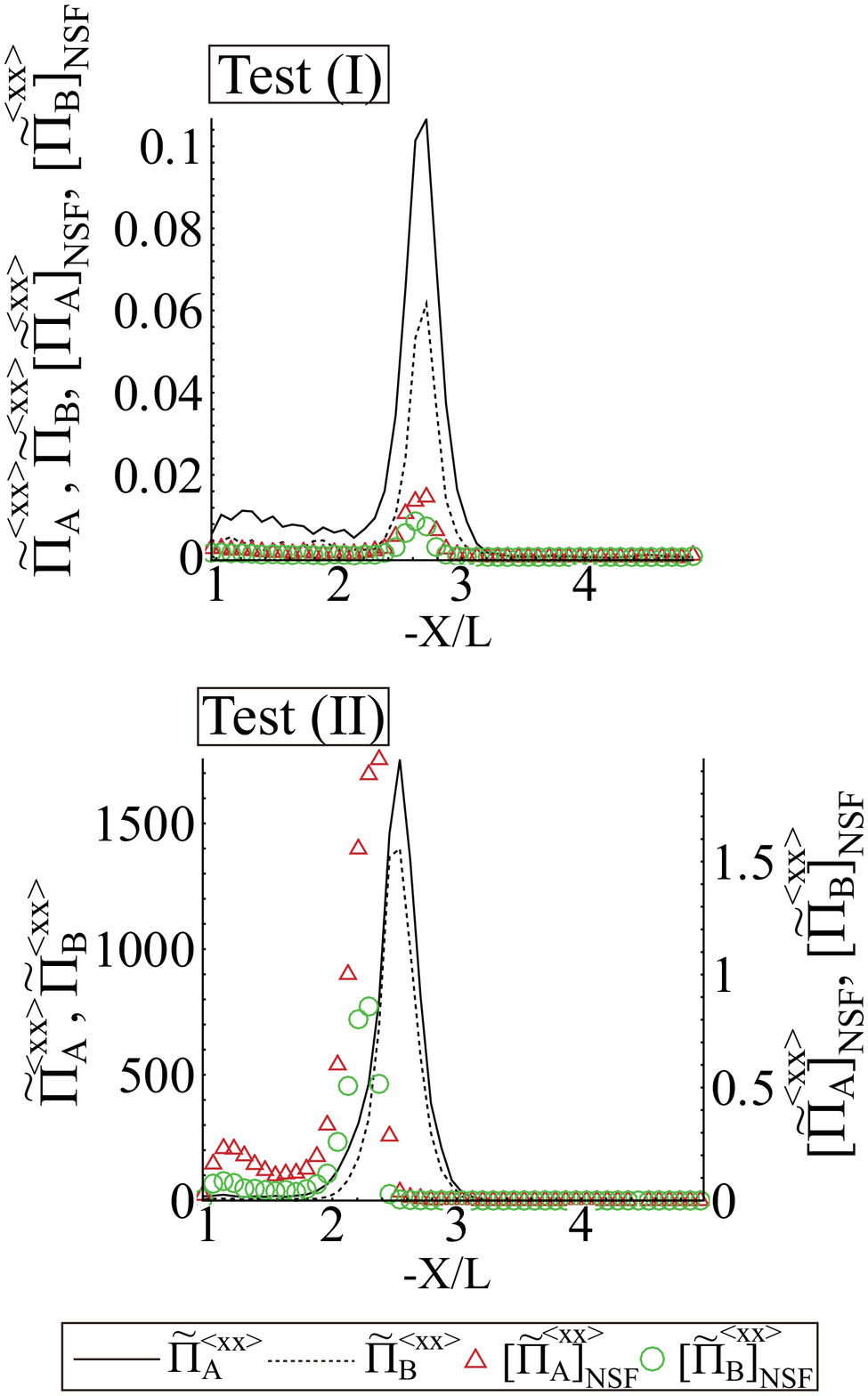}\\
\small{FIG. 5: Profiles of $\tilde{\Pi}_i^{\left<xx\right>}$ and $\left[\tilde{\Pi}_i^{\left<xx\right>}\right]_{\mbox{\tiny{NSF}}}$ ($i=$A, B) along the SSL in Tests (I) (upper frame) and (II) (lower frame).}
\end{center}
\begin{center}
\includegraphics[width=0.75\textwidth]{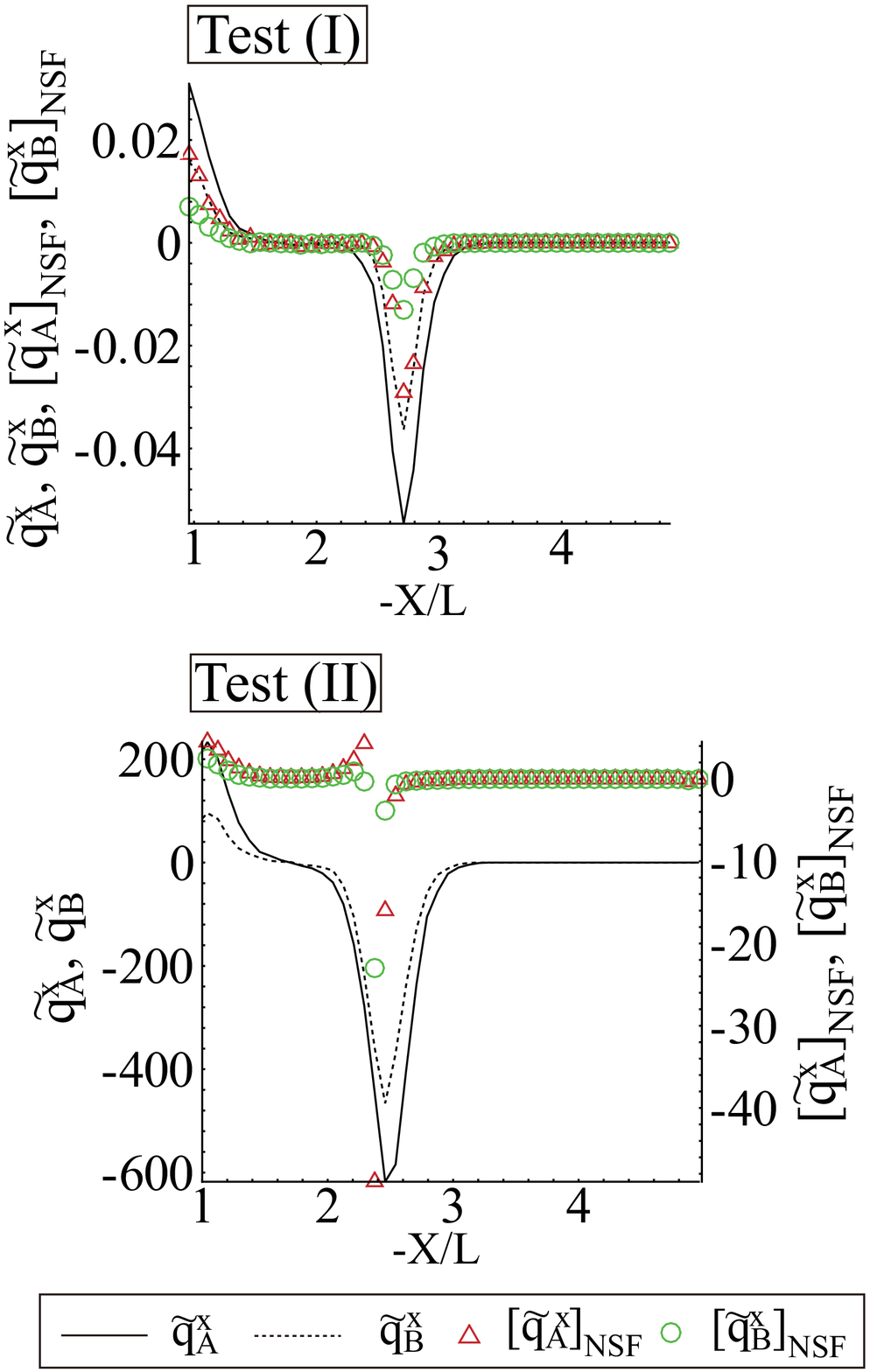}\\
\small{FIG. 6: Profiles of $\tilde{q}_i^{x}$ and $\left[\tilde{q}_i^{x}\right]_{\mbox{\tiny{NSF}}}$ ($i=$A, B) along the SSL in Tests (I) (upper frame) and (II) (lower frame).}
\end{center}
\begin{center}
\includegraphics[width=0.75\textwidth]{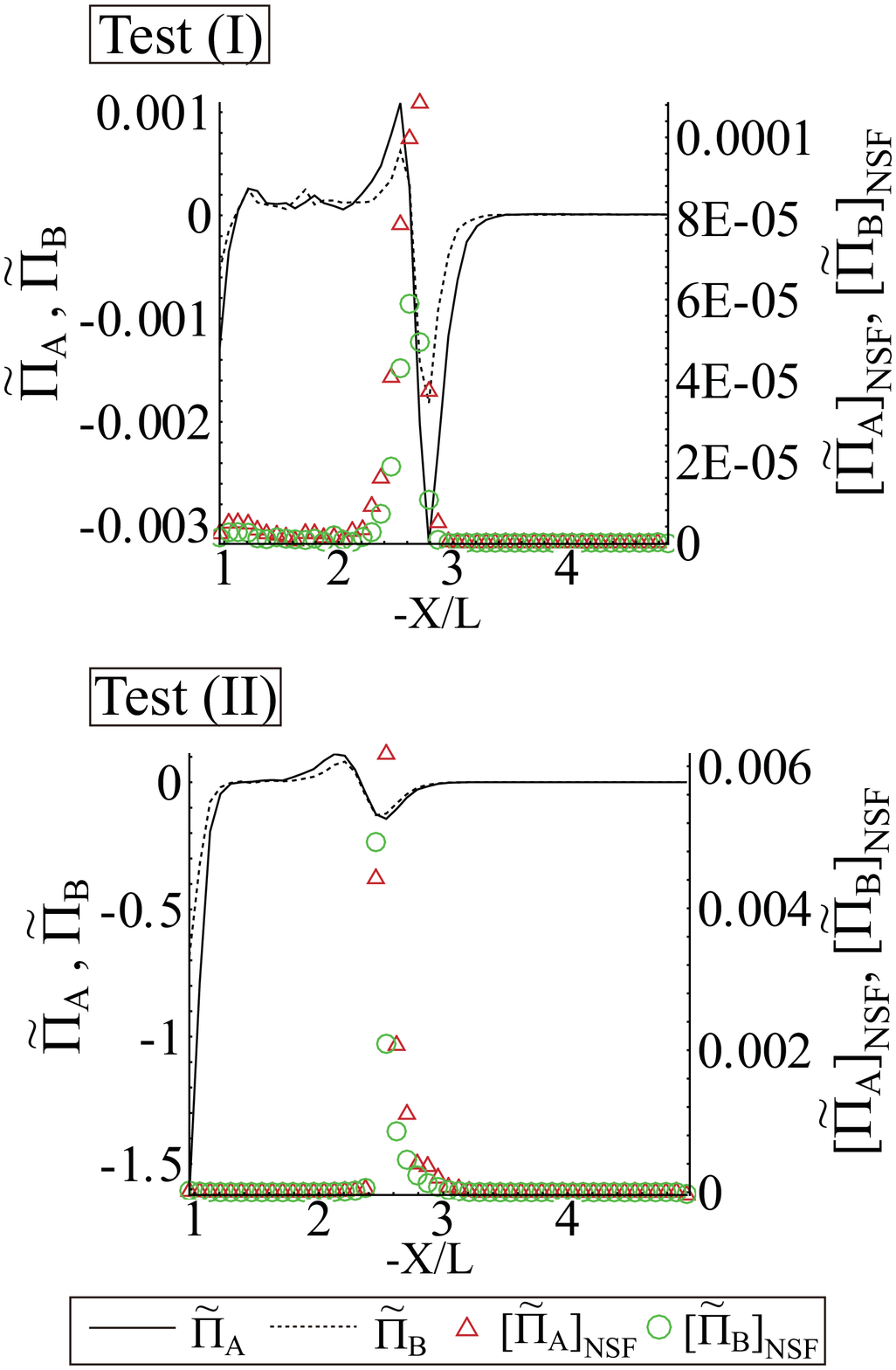}\\
\small{FIG. 7:Profiles of $\tilde{\Pi}_i$ and $\left[\tilde{\Pi}_i \right]_{\mbox{\tiny{NSF}}}$ ($i=$A, B) along the SSL in Tests (I) (upper frame) and (II) (lower frame).}
\end{center}
As mentioned above, the nonequilibrium between species A and B appears in the thermal boundary layer in Test (II), as shown in the lower frame of Fig. 2, whereas the nonequilibrium between species A and B seems to disappear behind the shock wave in Test (I), as shown in the upper frame of Fig. 2. Figure 8 shows distribution functions, $f_i\left(v^x_i\right)=\int \gamma\left(v_i\right)^5 f_i d v_i^y d v_i^z/n_i$ versus $\tilde{v}^x_i$ ($i=A,B$) on the SSL in Tests (I) (upper frame) and (II) (lower frame). Differences between $f_A\left(\tilde{v}^x_A\right)$ and $f_B\left(\tilde{v}_B^x\right)$ mean the nonequilibrium between species A and B. The upper frame of Fig. 8 shows that the nonequilibrium between species A and B exists at point (A) $-X/L=2.96$, which corresponds to the forward regime of the shock wave, as shown in the upper frame of Fig. 2, whereas the nonequilibrium between species A and B disappears at point (B) $-X/L=2.54$, which corresponds to the backward regime of the shock wave, as shown in the upper frame of Fig. 2. On the other hand, the lower frame of Fig. 8 shows that the nonequilibrium between species A and B exists at point (A) $-X/L=1.38$, which corresponds to the thermal boundary layer, as shown in the lower frame of Fig. 2, whereas the nonequilibrium between species A and B still exists at point (B) $-X/L=1.04$, which corresponds to the vicinity of the wall, as shown in the lower frame of Fig. 2. As a result, we can conclude that the nonequilibrium between species A and B is not dismissed behind the shock wave in Test (II).
\begin{center}
\includegraphics[width=0.5\textwidth]{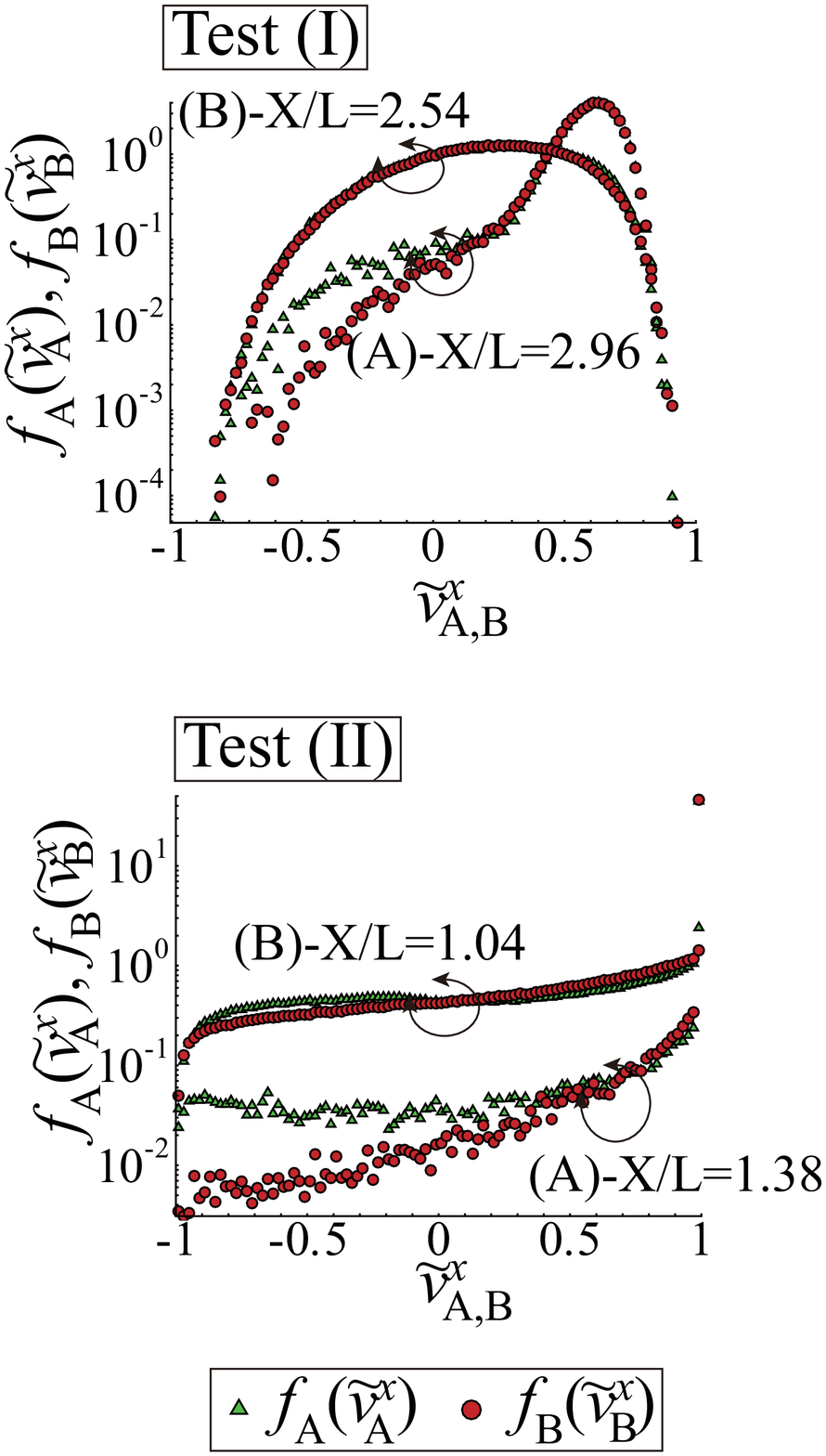}\\
\small{FIG. 8: $f_A\left(\tilde{v}^x_A\right)$ versus $\tilde{v}^x_A$ and $f_B\left(\tilde{v}_B^x\right)$ versus $\tilde{v}^x_B$ at points (A) ($-X/L=2.96$) and (B) ($-X/L=2.54$) on the SSL in Test (I) (upper frame), and $f_A\left(\tilde{v}^x_A\right)$ versus $\tilde{v}^x_A$ and $f_B\left(\tilde{v}^x_B\right)$ versus $\tilde{v}^x_B$ at points (A) ($-X/L=1.38$) and (B) ($-X/L=1.04$) on the SSL in Test (II) (lower frame).}
\end{center}
\subsection{Effect of mass ratio $m_A/m_B$ on dissipation process of binary mixture gas}
Finally, we numerically investigate effects of the mass ratio ($m_A/m_B$) on dissipation process using $m_A/m_B=0.25$ and $m_A/m_B=0.5$ together with $m_A/m_B=1$ in Test (II), when $\tilde{m}_B=1$, $\tilde{d}_A=0.5$ and $\tilde{d}_B=1$. Other conditions of the uniform flow are same as those in Test (II).\\
Figure 9 shows profiles of the flow velocity (upper frame) and temperature (lower frame) along the SSL, when $\tilde{m}_A=0.25$, $0.5$ and $1$. Figure 9 shows that the location of the shock wave moves toward the wall, as $\tilde{m}_A$ increases. The shock wave separation increases, as $\tilde{m}_A$ decreases. The increase of the shock wave separation yields the increase of the difference between $\tilde{U}_A^x$ and $\tilde{U}_B^x$ in Eq. (44). As a result, we can easily predict that the diffusion flux increases, as the shock wave separation increases. The overshoot of $\theta_A$ becomes larger, when $\tilde{m}_A$ decreases from $\tilde{m}_A=0.5$. $\theta_A$ is similar to $\theta_B$ behind the shock wave, when $\tilde{m}_A=0.25$ and $0.5$.
\begin{center}
\includegraphics[width=0.5\textwidth]{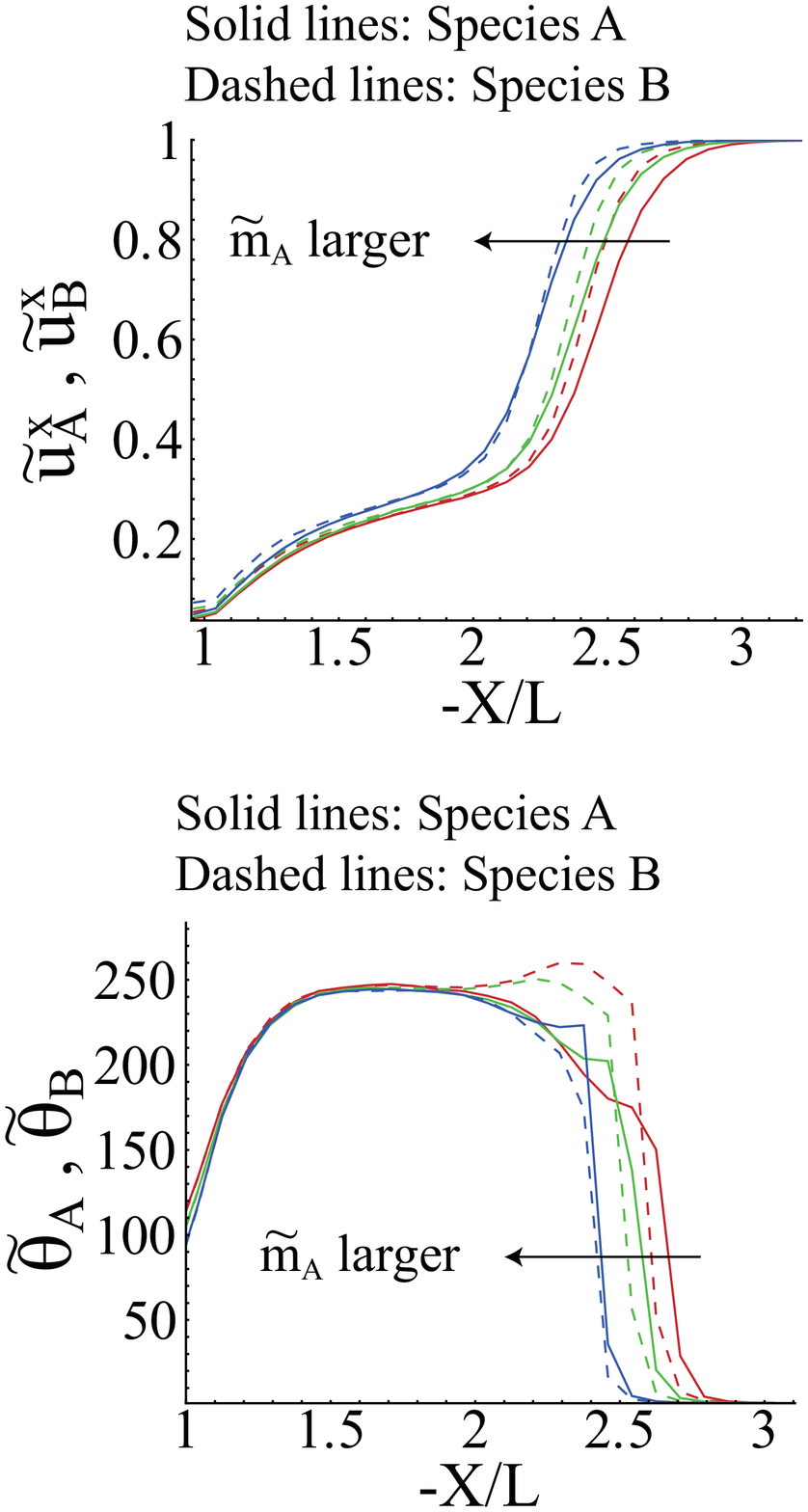}\\
\small{FIG. 9: Profiles of $\tilde{u}^x_i$ (upper frame) and $\tilde{\theta}_i$ (lower frame) ($i=$A, B) along the SSL, when $\tilde{m}_A=0.25$, $0.5$ and $1$.}
\end{center}
Figure 10 shows profiles of diffusion fluxes, $\tilde{J}_{AB}^x$, $\tilde{J}_{BA}^x$ and $\tilde{\mathfrak{J}}_{AB}^x$ along the SSL, when $\tilde{m}_A=0.5$ (upper frame) and $0.25$ (lower frame). Firstly, we obtain $\tilde{J}_{AB}^x=-\tilde{J}_{BA}^x$. The absolute value of the negative peak of $\tilde{J}_{AB}^x$ inside the shock wave increases, as $\tilde{m}_A$ decreases, owing to the increase of the shock separation. $\tilde{\mathfrak{J}}_{AB}^x$ is markedly similar to $\tilde{J}_{AB}^x$ in the range of $1 \le -X/L \le 2.5$, when $\tilde{m}_A=0.5$, whereas $\tilde{\mathfrak{J}}_{AB}^x$ is markedly similar to $\tilde{J}_{AB}^x$ in the range of $1 \le -X/L \le 2.4$, when $\tilde{m}_A=0.25$. Consequently, the diffusion flux does not depend on the mass ratio $m_A/m_B$.
\begin{center}
\includegraphics[width=0.5\textwidth]{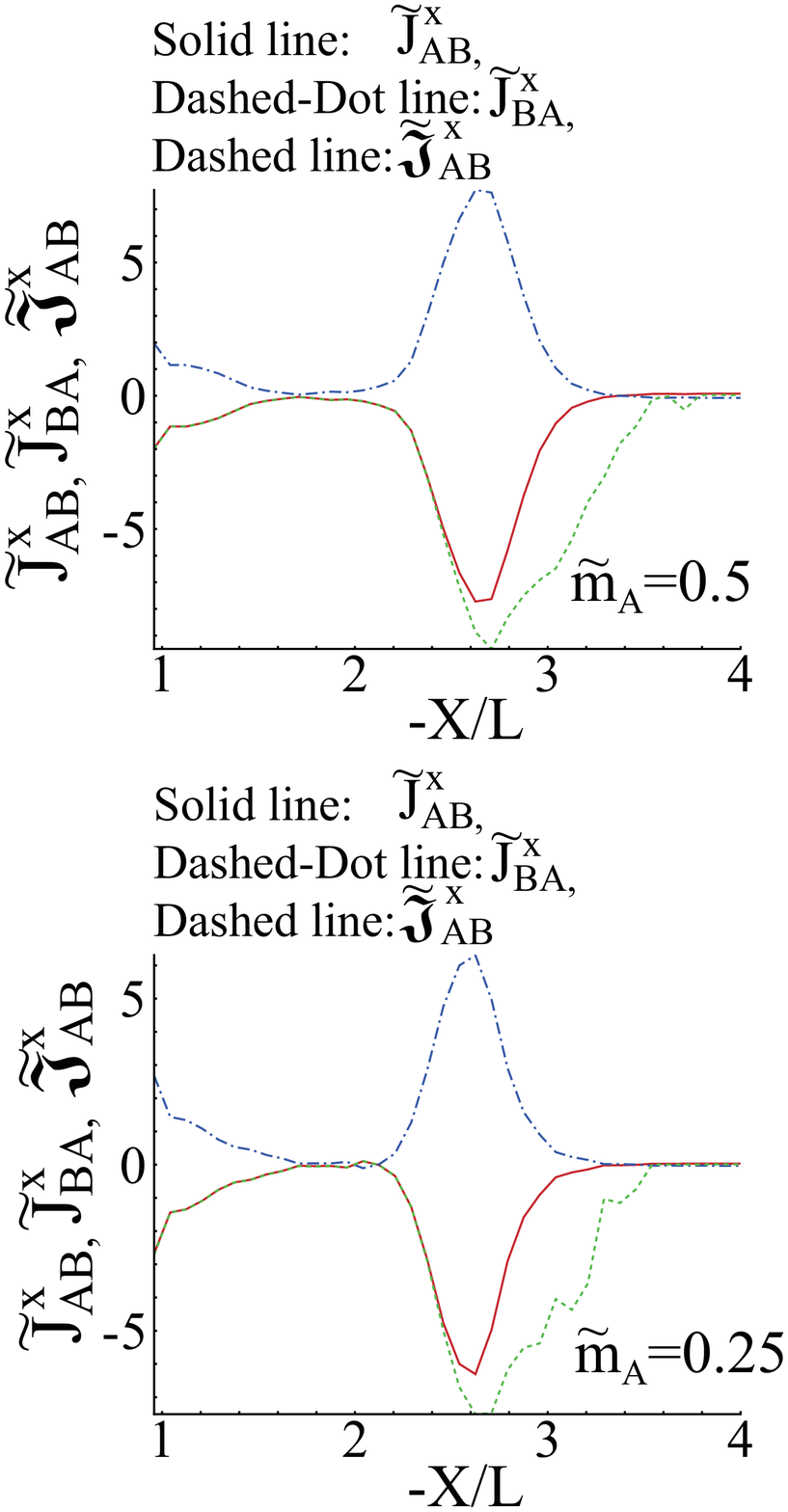}\\
\small{FIG. 10: Profiles of $\tilde{J}_{AB}^x$, $\tilde{J}_{BA}^x$ and $\tilde{\mathfrak{J}}_{AB}^x$ ($i=$A, B) along the SSL, when $\tilde{m}_A=0.5$ (upper frame) and $0.5$ (lower frame).}
\end{center}
Finally, we put further comments on the overshoot of $\theta_B$, when $\tilde{m}_A=0.25$ and $\tilde{m}_A=0.5$. The increase of the shock wave separation means the increase of the deceleration of the species $B$ in accordance with the deceleration of the species A inside the shock wave. The increase of the deceleration of $\tilde{u}_B^x$ yields an increase of the thermal energy of the species B, which is converted from its kinetic energy. Such an increase of the thermal energy leads to overshoots of $\tilde{\theta}_B$. Figure 11 shows profiles of $\tilde{\theta}_A$, $\tilde{\theta}_B$, and the averaged temperature $\bar{\theta}$ along the SSL, when $\tilde{m}_A=0.5$ (upper frame) and $\tilde{m}_A=0.25$ (lower frame). $\tilde{\theta}_B$ decreases toward $\bar{\theta}$ behind the point of its overshoot, when $\tilde{m}_A=0.25$ and $0.5$. $\tilde{\theta}_A$ increases toward $\bar{\theta}$ in the range of $1.77 \le -X/L \le 2.36$, when $\tilde{m}_{A}=0.5$, whereas $\tilde{\theta}_A$ increases toward $\bar{\theta}$ in the range of $1.87 \le -X/L \le 2.47$, when $\tilde{m}_{A}=0.25$. Of course, such relaxations of $\tilde{\theta}_A$ and $\tilde{\theta}_B$ to $\bar{\theta}$ are expressed by the term ${U_a}_\beta \Psi_a^\beta$ in Eq. (23).
\begin{center}
\includegraphics[width=0.5\textwidth]{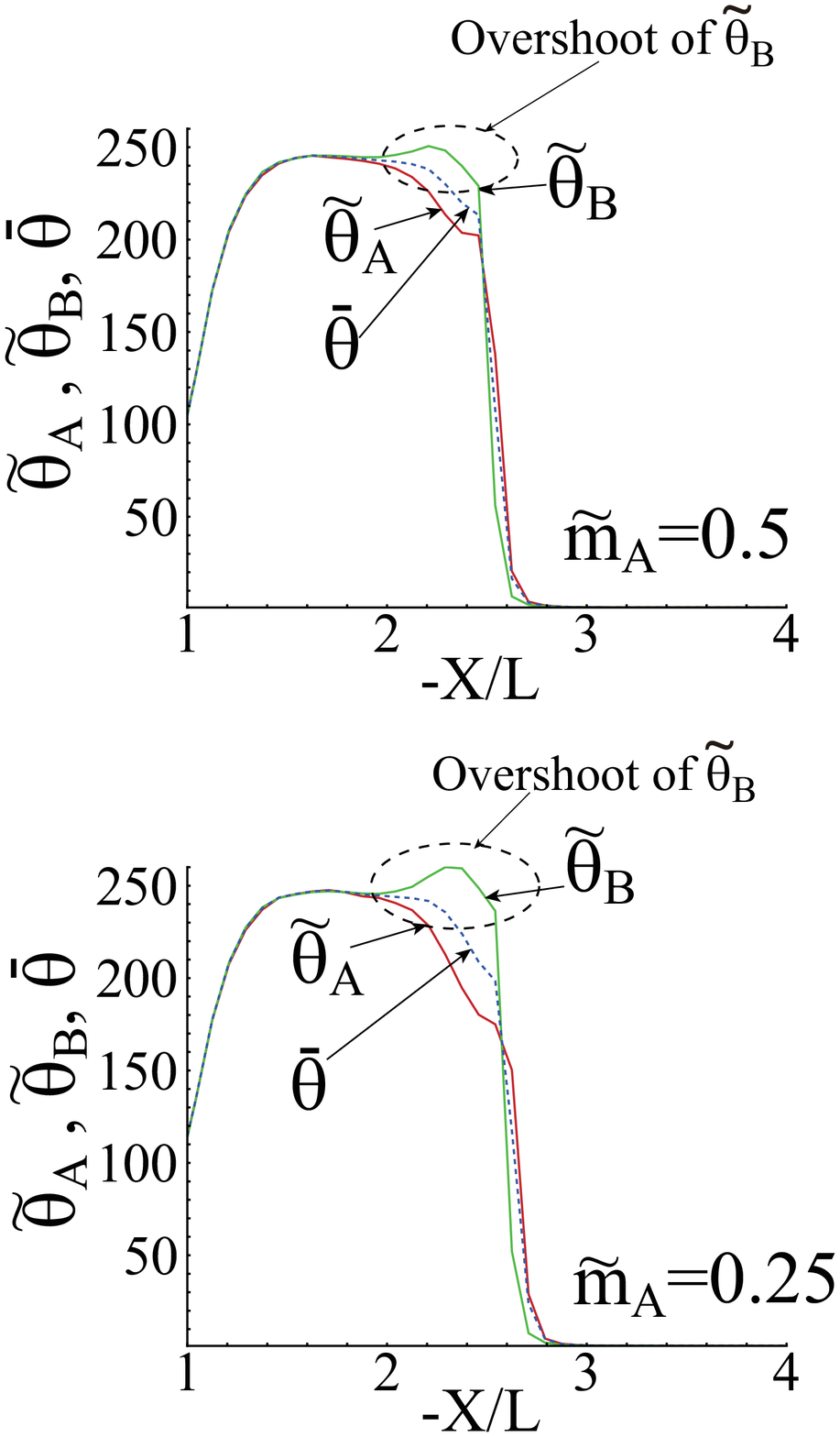}\\
\small{FIG. 11: Profiles of $\tilde{\theta}_A$, $\tilde{\theta}_B$ and $\bar{\theta}$ along the SSL, when $\tilde{m}_A=0.5$ (upper frame) and $\tilde{m}_A=0.25$ (lower frame).}
\end{center}
\section{Concluding Remarks}
In this paper, we investigated dissipation process of the thermally relativistic binary mixture gas, which is composed of hard spherical particles. \textcolor{black}{In particular, we investigated the characteristics of the diffusion flux for the thermally relativistic binary mixture gas by solving the RBE, numerically.} We used the diffusion and thermal-diffusion coefficients calculated by Kox \textit{et al}. to calculate the NSF order approximation of the diffusion flux. As an object of the numerical analysis, the thermally relativistic rarefied-shock layer of the binary mixture gas around the triangle prism was investigated by solving the RBE on the basis of the DSMC method. In the profile of the diffusion flux along the SSL, the diffusion flux via the thermal-diffusion coefficient is dominant over the diffusion flux via the diffusion coefficient inside the shock wave, whereas the difference between the first order approximation of the diffusion coefficient and the second order approximation of the diffusion coefficient by Kox \textit{et al}. is markedly small. The profile of the diffusion flux inside the shock wave, which is obtained using the DSMC method, is roughly approximated by the NSF order approximation on the basis of transport coefficients by Kox \textit{et al}., when the Lorenz contraction in the uniform flow is small. The NSF order approximation of the diffusion flux has a positive peak in the forward regime of the shock wave, where the diffusion flux, which is obtained using the DSMC method, is approximately zero, when the Lorenz contraction in the uniform flow is large. The negative peak of the diffusion flux inside the shock wave, which is obtained using the DSMC method, however, is roughly approximated by the NSF order approximation on the basis of transport coefficients by Kox \textit{et al}., although magnitudes of peak values of the dynamic pressure, pressure deviator and heat flux inside the shock wave, which are obtained using the DSMC method, are markedly larger than those approximated using the NSF law, because the generic Knudsen number becomes large. Then, we formulated the approximate diffusion flux ($\tilde{\mathfrak{J}}_a^x$) using the product of the difference between four velocities of two species of hard spherical particles with a half of averaged number density. $\tilde{\mathfrak{J}}_a^x$ is markedly similar to the diffusion flux, which is obtained using the DSMC method, when the local Lorentz contraction is not so large. Such a similarity confirms that the diffusion flux is surely independent of the generic Knudsen number from its definition. The nonequilibrium between two species of hard spherical particles remains behind the shock wave, when the Lorentz contraction in the uniform flow is large, whereas the nonequilibrium between two species of hard spherical particles disappears behind the shock wave, when the Lorentz contraction in the uniform flow is small. Such a remained nonequilibrium state in the vicinity of the wall yields differences between flow velocities of two species of hard spherical particles, which yields the diffusion flux, whose magnitude is markedly larger than the magnitude of the NSF order approximation, when the Lorentz contraction in the uniform flow is large. Thus, we consider that the relaxation process of the nonequilibrium state between two species strongly depends on the local Lorentz contraction together with the local temperature. Finally, we numerically investigated effects of the mass ratio between two species on dissipation process. $\tilde{\mathfrak{J}}_a^x$ is still markedly similar to the diffusion flux, which is obtained using the DSMC method, even when the mass ratio of two species of hard spherical particles is changed. The decrease of the mass ratio emphasizes the overshoot of the temperature of the heavier hard spherical particles, whereas the peak absolute value of the diffusion flux inside the shock wave increases in accordance with the decrease of the mass ratio. Such an overshoot of the temperature of the heavier hard spherical particles is explained by the increase of the shock wave separation in accordance with the decrease of the mass ratio, which leads to the increase of the magnitude of the diffusion flux inside the shock wave. \textcolor{black}{In summary, the diffusion flux must be calculated from the particle four flow, which is formulated using the four velocity distinguished by each species of particles, although we need not consider the diffusion flux, when Grad's moment equations are formulated using the four velocity distinguished by each species of particles. Finally, numerical results confirm that the Fourier law for the binary mixture gas, which neglects all the diffusive terms, approximates heat fluxes with worse accuracies than the Fourier law for the single component gas does, owing to neglect of all the diffusive terms.}
\begin{appendix}
\section{Comments on right hand sides of Eqs. (19)-(21)}
We mention to the right hand sides of Eqs. (19)-(21). As mentioned in Sec. II-(B), We used $f_a$ and $f_b$, which are approximated using Grad's 14 moment equations in Eq. (14), to evaluate $\Psi_a^\beta$ and $\Psi_a^{\beta\gamma}$ in Eqs. (4) and (5). We can easily conjecture that $\Psi_a^\beta$ and $\Psi_a^{\beta\gamma}$ are functions of Grad's 28 ($=2 (\mbox{number of species}) \times 14$) moments, when we consider the binary mixture gas, which is composed of two species (\enquote{$a$} and \enquote{$b$}) of hard spherical particles with equal masses.\\
Actually, collisional moments of $\Pi_a$, $\Pi_a^{\left<\alpha\beta\right>}$ and $q_a^\alpha$ were calculated by Kremer and Marquis Jr. for the binary mixture gas of Israel-Stewart particles with similar masses in the following forms, when the flow velocities and temperatures of two species are equal to each other: \cite{Kremer}
\begin{eqnarray}
&&\frac{C_2}{2}D \Pi_a+...=-\mathfrak{X}_{1,a} \Pi_a-\mathfrak{X}_{2,a}\left(\Pi_b-\Pi_a\right)+\mathfrak{F}_a,\\
&&C_4 D \Pi_a^{\left<\alpha\beta\right>}+...=-\mathfrak{X}_{3,a} \Pi_a^{\left<\alpha\beta\right>}-\mathfrak{X}_{4,a}\left(\Pi_b^{\left<\alpha\beta\right>}-\Pi_a^{\left<\alpha\beta\right>}\right),\\
&&5 C_3 D q_a^{\alpha}+...=-\mathfrak{X}_{5,i} q_a^\alpha-\mathfrak{X}_{6,a}\left(q_b^\alpha-q_a^\alpha\right)+\mathfrak{G}_a,
\end{eqnarray}
where $\mathscr{X}_{\ell,a}$ ($\ell=1,2,3,4,5,6$) are dissipation rates, and $\mathfrak{F}_a$ and $\mathfrak{G}_a$ are expressed using five field variables \cite{Kremer}. From Eqs. (A1)-(A3), we can obtain the NSF law by solving three sets of simultaneous equations, namely, simultaneous equations of $\Pi_a$ and $\Pi_b$, $\Pi_a^{\left<\alpha\beta\right>}$ and $\Pi_b^{\left<\alpha\beta\right>}$, and $q_a^\alpha$ and $q_b^\alpha$ using the first Maxwellian iteration. As a result, we can readily predict that the right hand sides of Eqs. (19)-(21) can be written in similar forms to those in Eqs. (A1)-(A3). Provided that the right hand side of Eq. (19) can be expressed with the linear combination of $\Pi_a$ and $\Pi_b$, the Navier-Stokes law for $\Pi_a$ must be written with the linear combination of $\nabla_{a \alpha} U_a^\alpha$ and $\nabla_{b \alpha} U_b^\alpha$ as a result of the first Maxwellian iteration of moment equations of $\Pi_a$ and $\Pi_b$. We, however, are unable to calculate $\mathfrak{X}_{\ell,i}$ in the case of hard spherical particles with equal masses. Then, we set $\mathfrak{X}_{1, a}=\sum_b \mathfrak{B}_{ab}$, $\mathfrak{X}_{3, a}=\sum_b \mathfrak{C}_{ab}$ and $\mathfrak{X}_{5, a}=\sum_b \mathfrak{D}_{ab}$ in right hand sides of Eqs. (19)-(21) . In Eq. (A1), the term with $\Pi_b-\Pi_a$ is equal to zero, when we consider the single component gas. Therefore, the term with $\Pi_b-\Pi_a$ is the diffusive term between $\Pi_a$ and $\Pi_b$, which is included in the diffusive term $\delta \Pi_{ab}$ in the right hand side of Eq. (19). In the first Maxwellian iteration of Eqs. (20), (25) and (26), we assume that $\left|\left[\Pi_b\right]_{\mbox{\tiny{NSF}}}-\left[\Pi_a\right]_{\mbox{\tiny{NSF}}}\right| \ll \left| \left[\Pi_a\right]_{\mbox{\tiny{NSF}}}\right|$, because we cannot evaluate $\mathfrak{X}_{\ell,a}$ ($\ell=1,2$) in right hand side of Eq. (25). Similarly, we assume that $\left|\left[\Pi_b^{\left<\alpha\beta\right>}\right]_{\mbox{\tiny{NSF}}}-\left[\Pi_a^{\left<\alpha\beta\right>}\right]_{\mbox{\tiny{NSF}}}\right| \ll \left|\left[\Pi_a^{\left<\alpha\beta\right>}\right]_{\mbox{\tiny{NSF}}}\right|$ and $\left|\left[q_b^{\alpha}\right]_{\mbox{\tiny{NSF}}}-\left[q_a^{\alpha}\right]_{\mbox{\tiny{NSF}}}\right| \ll \left|\left[q_a^\alpha \right]_{\mbox{\tiny{NSF}}}\right|$ in right hand sides of Eqs. (20) and (26). As a result of such assumptions, we obtain Eqs. (27)-(29) by setting terms with $\left[\Pi_b^{\left<\alpha\beta\right>}\right]_{\mbox{\tiny{NSF}}}-\left[\Pi_a^{\left<\alpha\beta\right>}\right]_{\mbox{\tiny{NSF}}}$, $\left[\Pi_b^{\left<\alpha\beta\right>}\right]_{\mbox{\tiny{NSF}}}-\left[\Pi_a^{\left<\alpha\beta\right>}\right]_{\mbox{\tiny{NSF}}}$ and $\left[q_b^{\alpha}\right]_{\mbox{\tiny{NSF}}}-\left[q_a^{\alpha}\right]_{\mbox{\tiny{NSF}}}$ as zero in $\delta \Pi_{ab}$, $\delta \Pi_{ab}^{\left<\alpha\beta\right>}$ and $\delta q_{ab}^\alpha$ in Eqs. (20), (25) and (26), when we take the first Maxwellian iteration of Eqs. (20), (25) and (26). Here, we remind that diffusive terms of in $\Pi_a$, $\Pi_a^{\left<\alpha\beta\right>}$ and $q_a^\alpha$ ($i=a,b$)  in $\Psi_a^\alpha$ are set as zero in the left hand sides of Eqs. (25) and (26), as discussed in Ref. \cite{memo3}.
\end{appendix}


\begin{thebibliography}{20}
\bibitem{memo3} In the single component relativistic gas, we obtain $\Psi_a^\alpha=0$. Then, we can easily conjecture that $\Psi_a^\alpha$ is expressed with a combination of $\Pi_b, \Pi_a, \Pi_b^{\left<\alpha\beta\right>}, \Pi_a^{\left<\alpha\beta\right>}, q_b^\alpha$ and $q_a^\alpha$ ($(a,b)=(A,B)$ or $(B,A)$) with a similar analogy, which is discussed in appendix A. Therefore, the terms with $\Pi_b, \Pi_a, \Pi_b^{\left<\alpha\beta\right>}, \Pi_a^{\left<\alpha\beta\right>}, q_b^\alpha$ and $q_a^\alpha$ can be transferred from the left to the right hand sides of Eqs. (25) and (26) as diffusive terms, because $\Pi_a$, $\Pi_a^{\left<\alpha\beta\right>}$ and $q_a^\alpha$ are not expressed with their derivative terms. We, however, assume that $\left[\Pi_i\right]_{\mbox{\tiny{NSF}}}=\left[\Pi_i^{\left<\alpha\beta\right>}\right]_{\mbox{\tiny{NSF}}}=\left[q_i^\alpha\right]_{\mbox{\tiny{NSF}}}=0$ ($i=a,b$) in $\Psi_a^\alpha$ by neglecting such diffusive terms with $\Pi_b, \Pi_a, \Pi_b^{\left<\alpha\beta\right>}, \Pi_a^{\left<\alpha\beta\right>}, q_b^\alpha$ and $q_a^\alpha$ in left hand sides of Eqs. (25) and (26).
\bibitem{memo} Readers must notice that two types of nonequilibrium states exist in this problem, when $m_A=m_B$. One is the nonequilibrium in each species A and B, in which $f_A \neq f_A^{MJ}$ and $f_B \neq f_B^{MJ}$ are allowable, even when $f_A=f_B$. The other is the nonequilibrium between species A and B, in which $f_A= f_A^{MJ}$, $f_B=f_B^{MJ}$ and $f_A \neq f_B$ are allowable. Therefore, the perfect equilibrium state is realized by $f_A=f_B=f_A^{MJ}=f_B^{MJ}$.
\end{thebibliography}
\end{document}